\documentclass[twoside,twocolumn,9pt]{article}
\usepackage{extsizes}
\usepackage[super,sort&compress,comma]{natbib} 
\usepackage[version=3]{mhchem}
\usepackage[left=1.5cm, right=1.5cm, top=1.785cm, bottom=2.0cm]{geometry}
\usepackage{balance}
\usepackage{mathptmx}
\usepackage{sectsty}
\usepackage{graphicx} 
\usepackage{lastpage}
\usepackage[format=plain,justification=justified,singlelinecheck=false,font={stretch=1.125,small,sf},labelfont=bf,labelsep=space]{caption}
\usepackage{float}
\usepackage{fancyhdr}
\usepackage{fnpos}
\usepackage[english]{babel}
\addto{\captionsenglish}{%
  
}
\usepackage{array}
\usepackage{droidsans}
\usepackage{charter}
\usepackage[T1]{fontenc}
\usepackage[usenames,dvipsnames]{xcolor}
\usepackage{setspace}
\usepackage[compact]{titlesec}
\usepackage{hyperref}

\definecolor{cream}{RGB}{222,217,201}

\usepackage{amsmath,amssymb}
\def\kT{\ensuremath{k_\text{B}T}}

\DeclareMathOperator\smol{\Omega}
\DeclareMathOperator\dsmol{\delta\Omega}
\DeclareMathOperator\projP{\mathcal P}
\DeclareMathOperator\projQ{\mathcal Q}
\newcommand{\tlname}[1]{\ensuremath{\mathit{#1}}}
\newcommand{\Pe}{\tlname{Pe}}
\let\bs\boldsymbol

\usepackage{glossaries}
\makeatletter
\newcommand*{\glsplainhyperlink}[2]{%
  \colorlet{currenttext}{.}
  \colorlet{currentlink}{\@linkcolor}
  \hypersetup{linkcolor=currenttext}
  \hyperlink{#1}{#2}%
  \hypersetup{linkcolor=currentlink}
}
\let\@glslink\glsplainhyperlink
\makeatother
\newacronym{ABP}{ABP}{active Brownian particle}
\newacronym{MCT}{MCT}{mode-coupling theory of the glass transition}
\newacronym{ABPMCT}{ABP-MCT}{mode-coupling theory for active Brownian particles}
\newacronym{ITT}{ITT}{integration-through transients}
\newacronym{MSD}{MSD}{mean-squared displacement}
\newacronym{EDBD}{ED-BD}{event-driven Brownian dynamics}
\newacronym{DFT}{DFT}{density-functional theory}
\newacronym{MHNC}{MHNC}{modified hypernetted-chain}
\newacronym{AOUP}{AOUP}{active Ornstein-Uhlenbeck particles}
\newacronym{2D}{2D}{two-dimensional}
\newacronym{3D}{3D}{three-dimensional}

\hypersetup{colorlinks=true, linkcolor=BrickRed, urlcolor=blue!50!black, citecolor=blue!50!black}

\begin{document}
\pagestyle{fancy}
\thispagestyle{plain}
\fancypagestyle{plain}{

\renewcommand{\headrulewidth}{0pt}
}

\makeFNbottom
\makeatletter
\renewcommand\LARGE{\@setfontsize\LARGE{15pt}{17}}
\renewcommand\Large{\@setfontsize\Large{12pt}{14}}
\renewcommand\large{\@setfontsize\large{10pt}{12}}
\renewcommand\footnotesize{\@setfontsize\footnotesize{7pt}{10}}
\makeatother

\renewcommand{\thefootnote}{\fnsymbol{footnote}}
\renewcommand\footnoterule{\vspace*{1pt}%
\color{cream}\hrule width 3.5in height 0.4pt \color{black}\vspace*{5pt}} 
\setcounter{secnumdepth}{5}

\makeatletter 
\renewcommand\@biblabel[1]{#1}            
\renewcommand\@makefntext[1]%
{\noindent\makebox[0pt][r]{\@thefnmark\,}#1}
\makeatother 
\renewcommand{\figurename}{\small{Fig.}~}
\sectionfont{\sffamily\Large}
\subsectionfont{\normalsize}
\subsubsectionfont{\bf}
\setstretch{1.125} 
\setlength{\skip\footins}{0.8cm}
\setlength{\footnotesep}{0.25cm}
\setlength{\jot}{10pt}
\titlespacing*{\section}{0pt}{4pt}{4pt}
\titlespacing*{\subsection}{0pt}{15pt}{1pt}

\fancyfoot{}
\fancyfoot[LO,RE]{\vspace{-7.1pt}\includegraphics[height=9pt]{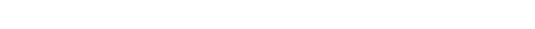}}
\fancyfoot[CO]{\vspace{-7.1pt}\hspace{13.2cm}\includegraphics{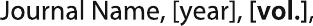}}
\fancyfoot[CE]{\vspace{-7.2pt}\hspace{-14.2cm}\includegraphics{head_foot/RF}}
\fancyfoot[RO]{\footnotesize{\sffamily{1--\pageref{LastPage} ~\textbar  \hspace{2pt}\thepage}}}
\fancyfoot[LE]{\footnotesize{\sffamily{\thepage~\textbar\hspace{3.45cm} 1--\pageref{LastPage}}}}
\fancyhead{}
\renewcommand{\headrulewidth}{0pt} 
\renewcommand{\footrulewidth}{0pt}
\setlength{\arrayrulewidth}{1pt}
\setlength{\columnsep}{6.5mm}
\setlength\bibsep{1pt}

\makeatletter 
\newlength{\figrulesep} 
\setlength{\figrulesep}{0.5\textfloatsep} 

\newcommand{\topfigrule}{\vspace*{-1pt}%
\noindent{\color{cream}\rule[-\figrulesep]{\columnwidth}{1.5pt}} }

\newcommand{\botfigrule}{\vspace*{-2pt}%
\noindent{\color{cream}\rule[\figrulesep]{\columnwidth}{1.5pt}} }

\newcommand{\dblfigrule}{\vspace*{-1pt}%
\noindent{\color{cream}\rule[-\figrulesep]{\textwidth}{1.5pt}} }

\makeatother

\twocolumn[
  \begin{@twocolumnfalse}
{\includegraphics[height=30pt]{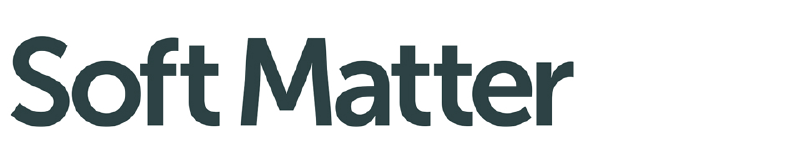}\hfill\raisebox{0pt}[0pt][0pt]{\includegraphics[height=55pt]{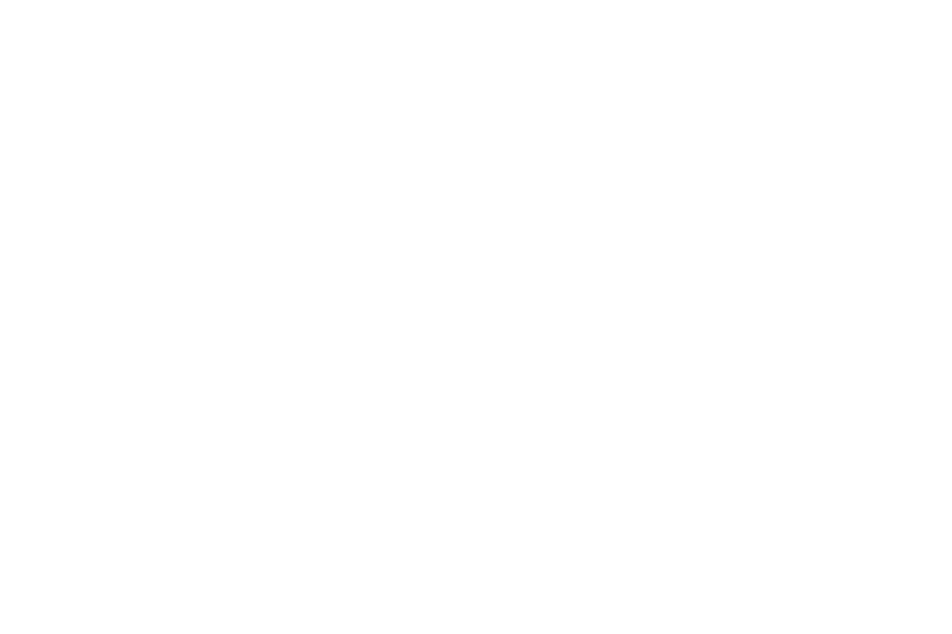}}\\[1ex]
\includegraphics[width=18.5cm]{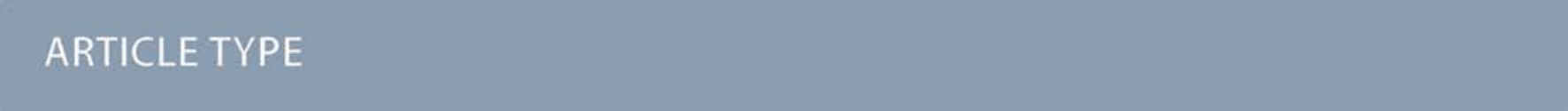}}\par
\vspace{1em}
\sffamily
\begin{tabular}{m{4.5cm} p{13.5cm} }

\includegraphics{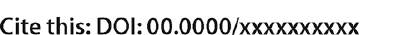} & \noindent\LARGE{\textbf{%
Tracer Dynamics in Crowded Active-Particle Suspensions
}}\\
\vspace{0.3cm} & \vspace{0.3cm} \\

 & \noindent\large{%
Julian Reichert$^{\ast}$\textit{$^{a}$}
and
Thomas Voigtmann\textit{$^{a,b}$}
}\\

\includegraphics{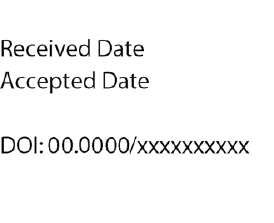} & \noindent\normalsize{%
We discuss the dynamics of \glspl{ABP} in crowded environments through
the \gls{MSD} of active and passive tracer particles in both active and
passive host systems.
Exact equations for the \gls{MSD} are derived using a projection operator
technique, extending the known solution for a single \gls{ABP}
to dense systems.
The interaction of the tracer particle with the host particles
gives rise to strong memory effects. Evaluating these approximately in
the framework of a recently developed \gls{ABPMCT}, we discuss the
various dynamical regimes that emerge: While self-propelled motion gives
rise to super-diffusive \gls{MSD}, at high densities, this competes with
an interaction-induced sub-diffusive regime. The predictions of the theory are
shown to be in good agreement with results obtained from an \gls{EDBD}
simulation scheme for the dynamics of two-dimensional
active Brownian hard disks.
}%
\end{tabular}

 \end{@twocolumnfalse} \vspace{0.6cm}

  ]

\renewcommand*\rmdefault{bch}\normalfont\upshape
\rmfamily
\section*{}
\vspace{-1cm}


\footnotetext{%
\textit{$^{a}$~Institut f\"ur Theoretische Physik, Universit\"at Innsbruck, Austria}}
\footnotetext{%
\textit{$^{b}$~Institut f\"ur Materialphysik im Weltraum, Deutsches Zentrum f\"ur Luft- und Raumfahrt (DLR), 51170 K\"oln, Germany}}
\footnotetext{%
\textit{$^{c}$~Department of Physics, Heinrich-Heine Universit\"at D\"usseldorf, Universit\"atsstr.~1, 40225 D\"usseldorf, Germany}}



\glsresetall
\section{Introduction}

The observation of active motion of self-propelled micro-organisms and the
peculiar collective effects that it gives rise to, is a fascinating topic
of biophysics that has stipulated a vast, rapidly growing research field
in soft matter and non-equilibrium statistical physics \cite{Elgeti.2015,Ramaswamy.2017}.
Since microswimmers are subject to both passive Brownian motion as well as
to active driving, the \gls{ABP} model \cite{Howse.2007}
has emerged as a convenient model system to study the interplay
of the two kinds of forces.
Interest in the \gls{ABP} model is further stirred by the experimental
realization through colloidal Janus particles \cite{Bechinger.2016}.
Direct observation in quasi-\gls{2D} setups ameks the \gls{MSD}
a key quantity to discuss \cite{Lozano.2019}.
The \gls{MSD} of a single \gls{ABP} can be obtained analytically,
and this already displays some interesting features: After a short-time
asymptote that is passive-diffusive, $\sim t$, a cross-over to
a super-diffusive transient, $\sim t^2$, signals persistent swimming,
before Brownian rotational diffusion leads to a long-time asymptote that
is again diffusive, but with an activity-enhanced diffusion coefficient.
The appearance of a super-diffusive regime in the \gls{MSD} also advocates
the non-equilibrium nature of the motion.

Analytical solutions for interacting \gls{ABP} are not available.
Yet the motion of both active and passive colloids in a crowded host
suspension (passive or active) is of high interest.
Recent observations revealed how a single Janus colloid changes its dynamics
when embedded in a glass-forming suspension of passive colloids
\cite{Lozano.2019}, or how colloidal motion is influenced by bacterial
baths \cite{Ortlieb.2019}.
The latter case, that of a passive colloid in an
active fluid, is a widely used micro-rheology technique to infer properties
of the host medium through the tracer-particle motion
\cite{Wirtz.2009}.
The \gls{MSD} of a passive tracer in active suspension shows clear
signs of the non-equilibrium bath dynamics
\cite{Wu.2000,Caspi.2000,Chen.2007,Wilhelm.2008,Gal.2010,Valeriani.2011,Lagarde.2020}.

Here we derive evolution equations to describe the \gls{MSD} of \gls{ABP}
in dense systems.
Using a Mori-Zwanzig projection-operator approach for the angle-resolved
tagged-particle density correlation function, we obtain in the limit
$q\to0$ two coupled integro-differential equations that contain the
coupling to the dense host system in three memory kernels corresponding
to translational (angular mode $l=0$) and dipolar-like ($l=\pm1$) couplings.
The equations reduce to ordinary differential equations whose solution
is the well known analytical result for a single \gls{ABP}, if these memory
kernels are dropped.
Modeling the memory kernels by a recent extension of the mode-coupling theory
of the glass transition, the \gls{ABPMCT} \cite{Liluashvili.2017,Reichert.2021,taggedpaper}, we obtain theoretical
predictions for the \gls{MSD} in the various cases of active and passive
tracers in active and passive host systems, and we compare those results
to \gls{EDBD} computer simulation.
As one nears kinetic arrest in the host system,
there emerges an interplay between the time scales of the free active
motion, and that of steric-hindrance induced caging provided by the
host particles.

\section{Theory}\label{sec:theory}

We consider the \acrfull{ABP} equations of motion
\begin{subequations}\label{eq:abpsde}
\begin{align}
  d\vec r_j&=\mu\vec F_j\,dt+\sqrt{2D_t}\,d\vec W_j+v_0\vec n(\varphi_j)\,dt
  \,,\\
  d\varphi_j&=\sqrt{2D_r}\,dW_{\varphi_j}\,.
\end{align}
\end{subequations}
The orientation of the ABP, $\vec n_j=\vec n(\varphi_j)=(\cos\varphi_j,\sin\varphi_j)^T$ evolves purely through rotational diffusion, where translational and rotational diffusion are driven by independent Wiener processes $d\vec W_j$ and
$d\vec W_{\varphi_j}$.
The $\vec F_j=-\nabla_jU$ are potential interaction forces that are taken to be steeply repulsive to model hard-disk behavior.
In absence of active driving, the system is in thermal equilibrium, thus the mobility obeys the fluctuation-dissipation theorem, $\kT\mu=D_t$.
In the active system, each particle experiences a constant self-propulsion force given by the swimming speed $v_0$, along its current director $\vec n_j$.

Into the $N$-particle system of \gls{ABP} with self-propulsion velocity
$v_0$, we embed a single tracer particle (position $\vec r_s$, orientation
$\varphi_s$) with self-propulsion velocity $v_0^s$, whose equations of
motion are again given by the equivalent of Eq.~\eqref{eq:abpsde}.
We allow for the case of a
tracer of different interactions than among the host particles,
and also different short-time diffusion coefficients $D_t^s$ and $D_r^s$,
although  in the discussion we will focus on otherwise identical particles
that merely differ in their self-propulsion speeds.
In particular this covers the experimentally relevant cases of a
passive tracer in an active host system ($v_0^s=0$, $v_0\neq0$) and
the reversed case of a single \gls{ABP} that is embedded in a
passive glass-forming fluid ($v_0^s\neq0$ but $v_0=0$) \cite{Lozano.2019}.

The \acrfull{MSD} is defined by
\begin{equation}
  \delta r^2(t)=\left\langle|\vec r_s(t)-\vec r_s(0)|^2\right\rangle\,,
\end{equation}
where $\langle\cdot\rangle$ denotes the ensemble average over realizations
of the \gls{ABP} system. We distinguish two important cases of averages:
that of the stationary non-equilibrium active system, leading to the
\emph{stationary} \gls{MSD} that we obtain from computer simulation.
For reasons that become apparent below, of theoretical interest is also
the \emph{transient} \gls{MSD} which is obtained by averaging over the
passive-equilibrium ensemble, keeping the active driving term in the
time evolution.
A tacit assumption made in comparing theory with simulation is that in
the parameter regime discussed below the different averages do not yield
qualitatively different results. This is corroborated by direct comparison
of stationary and transient averages in computer simulation \cite{taggedpaper}.
In the following derivation of the theory, $\langle\cdot\rangle$
denotes the \emph{equilibrium} distribution function of the corresponding
passive system.

\subsection{Transient correlation functions}

Equations~\eqref{eq:abpsde} describe a Markov process with configuration space elements $\Gamma=(\{\vec r_j\},\{\varphi_j\})$ whose probability distribution $p(\Gamma,t)$ evolves through the Smoluchowski equation $\partial_tp=\smol p$, with the Smoluchowski operator
\begin{equation}\label{eq:abpsmol}
  \smol=\sum_{j=1}^ND_t\vec\nabla_j\cdot\left(\vec\nabla_j-\beta\vec F_j\right)
  +D_r\partial_{\varphi_j}^2-v_0\vec\nabla_j\cdot\vec n_j\,.
\end{equation}
This operator consists of three parts that for later convenience we will
split according to $\smol(D_t,D_r,v_0)=\smol_\text{eq}(D_t,D_r)+\dsmol(v_0)
=\smol_T(D_t,v_0)+\smol_R(D_r)$ depending on context.
Here, $\smol_\text{eq}$ is the well-known equilibrium Smoluchowski operator whose stationary distribution $p_\text{eq}\propto\exp[-\beta U]$ defines the equilibrium averages of observables with the inverse temperature $\beta$.
The time evolution of the tracer particle is driven by the equivalent Smoluchowski
operator  including the tracer index in the particle sum, allowing for its
possibly distinct parameters $(D_t^s,D_r^s,v_0^s)$.

Microscopic observables of interest to describe the motion of the ABP are the
angle-resolved fluctuating particle densities
to wave vector $\vec q$ (of magnitude $q=|\vec q|$) and angular-mode
index $l$,
\begin{equation}
  \varrho_l(\vec q)=\sum_{k=1}^Ne^{i\vec q\cdot\vec r_k}e^{il\varphi_k}/\sqrt N\,,
\end{equation}
and their counterpart for the tracer particle,
\begin{equation}
  \varrho_l^s(\vec q)=e^{i\vec q\cdot\vec r^s}e^{il\varphi^s}\,,
\end{equation}
where the particle at $(\vec r^s,\varphi^s)$ is understood to be excluded from
the sum over the $N$ particles comprising the host system.
We will use the shorthand notation $\varrho_1\equiv\varrho_{l_1}(\vec q_1)$ where convenient.

The \gls{ABPMCT} is built on the \gls{ITT} framework \cite{Fuchs.2009} that allows to
treat formally the dynamical evolution of a non-equilibrium system with
arbitrarily strong perturbation, and provides a starting point for
approximations. In this framework, quantities of particular
interest are the transient
dynamical density correlation functions. In a spatially homogeneous system
they are diagonal in $\vec q$ and read
\begin{equation}
  \Phi_{ll'}(\vec q,t)=\left\langle\varrho_l^*(\vec q)\exp[\smol^\dagger t]\varrho_{l'}(\vec q)\right\rangle\,,
\end{equation}
where $\smol^\dagger$ is the adjoint, or backward, Smoluchowski operator,
\begin{equation}
  \smol^\dagger=\sum_{j=1}^ND_t\left(\vec\nabla_j+\beta\vec F_j\right)
  \cdot\vec\nabla_j+D_r\partial_{\varphi_j}^2+v_0\vec n_j\cdot\vec\nabla_j\,.
\end{equation}
The correlation function obeys $\Phi_{ll'}(\vec q,0)=S_{ll'}(q)$, where in the particular case of particles that interact through a spherically symmetric interaction potential,
\begin{equation}
  S_{ll'}(q)=\left\langle\varrho_l^*(\vec q)\varrho_{l'}(\vec q)\right\rangle
  =\delta_{ll'}\left(S(q)\delta_{l0}+(1-\delta_{l0})\right)\,.
\end{equation}
Here, $S(q)$ is the ordinary equilibrium static structure factor known from liquid state theory for the pasive system.

In a system that remains statistically isotropic,
the dynamical correlation functions obey specific transformation rules under
rotation,
\begin{equation}\label{eq:phitransform}
\bs\Phi(\vec q^{\,\prime},t)=\bs u^\dagger\cdot\bs\Phi(\vec q,t)\cdot\bs u\,,
\end{equation}
if $\vec q^{\,\prime}$ is the vector obtained by rotating $\vec q$ by an angle $\psi$, and $u_{ll'}=\delta_{ll'}\exp[il\psi]$.
In particular, letting $\theta_q$ be the angle of $\vec q$ with the $x$-axis normal $\vec e_x$,
\begin{equation}
  \tilde\Phi_{ll'}(q,t)=\Phi_{ll'}(q\vec e_x,t)=e^{i(l-l')\theta_q}\Phi_{ll'}(\vec q,t)\,.
\end{equation}
It follows that the diagonal elements of $\bs\Phi(\vec q,t)$ are isotropic even functions of $\vec q$.
In particular the positional density-correlation function $\Phi_{00}(q,t)$ is isotropic, even in $q$ and real-valued.

The tracer motion is characterized by the tagged-particle correlation
function,
\begin{equation}
  \phi^s_{ll'}(\vec q,t)=\left\langle\varrho_l^{s*}(\vec q)
  \exp[\smol^\dagger t]\varrho_{l'}^s(\vec q)\right\rangle\,,
\end{equation}
which obeys $\phi^s_{ll'}(\vec q,0)=\delta_{ll'}$.
Of particular interest here is the $(00)$ element of that correlation
function in the limit $q\to0$: it is by the rotation-transformation
property an isotropic, real-valued even function of $q$ and
linked to the \gls{MSD},
\begin{equation}
  \phi^s_{00}(q,t)=1-\frac{q^2}{4}\delta r^2(t)+\mathcal O(q^4)\,,
\end{equation}
in two spatial dimensions.

\subsection{Mori-Zwanzig evolution equations}

An exact evolution equation for the density-correlation functions can be
obtained through a projection operator scheme.
In essence, one rewrites the propagator
$\exp[\smol^\dagger t]$ in terms of its action on the projected subspace
$\projP$ spanned by the fluctuating densities, and a reminder that
remains within the orthogonal projection and gives rise to a non-Markovian
evolution of the projected variables expressed through memory integrals.
After taking care to transform the memory kernels into a form that
is one-particle irreducible, one obtains \cite{Liluashvili.2017,taggedpaper}
\begin{multline}\label{eq:mz}
  \partial_t\tilde{\bs\Phi}(q,t)
  +\tilde{\bs\omega}(q)\cdot\bs S^{-1}(q)\cdot\tilde{\bs\Phi}(q,t)
  \\
  +\int_0^tdt'\,\tilde{\bs m}(q,t-t')\cdot\left(\bs1\partial_{t'}+\tilde{\bs\omega}_R\right)\cdot\tilde{\bs\Phi}(q,t')=\bs 0\,
\end{multline}
where we have written $\tilde{\bs m}(q,t)=\tilde{\bs M}(q,t)\cdot
\tilde{\bs\omega}_T^{-1}(q)$.
An important quantity here and in the following is the frequency matrix
\begin{equation}
  \omega_{ll'}(\vec q)=-\left\langle\varrho_l^*(\vec q)\smol^\dagger\varrho_{l'}(\vec q)\right\rangle\,.
\end{equation}
It is decomposed into its rotational and translational parts,
$\bs\omega(\vec q)=\bs\omega_T(\vec q)+\bs\omega_R$, where
$\omega_{R,ll'}=\delta_{ll'}l^2D_r$ and $\bs\omega_T(\vec q)$ is given by
the tri-diagonal matrix
\begin{equation}\label{eq:omega}
  \tilde\omega_{ll'}(q)=\delta_{ll'}q^2D_t
  -\delta_{|l-l'|,1}\frac{iqv_0}{2}S_{ll}(q)\,.
\end{equation}

An analogous equation can be derived for the tagged-particle
correlator,
\begin{multline}\label{eq:mzs}
  \partial_t\tilde{\bs\phi}^s(q,t)
  +\tilde{\bs\omega}^s(q)\cdot\tilde{\bs\phi}^s(q,t)
  \\
  +\int_0^tdt'\,\tilde{\bs m}^s(q,t-t')\cdot\left(\bs1\partial_{t'}
  +\tilde{\bs\omega}_R^s\right)\cdot\tilde{\bs\phi}^s(q,t')=\bs0\,.
\end{multline}
Here, $\bs\omega^s(\vec q)$ is the analog of Eq.~\eqref{eq:omega} for
the tagged particle (for which formally $S_{ll}(q)=1$).

Equations \eqref{eq:mz} and \eqref{eq:mzs} constitute the starting point of
our discussion of the \gls{MSD} of active Brownian particles.
The memory kernels are given by formally exact expressions, e.g.,
\begin{equation}
  M^s_{ll'}(\vec q,t)=\left\langle\varrho^s_l(\vec q)\smol^\dagger_T
  \projQ\exp[\smol^\dagger_\text{irr}t]\projQ\smol^\dagger_T\varrho^s_{l'}
  \right\rangle\,,
\end{equation}
where $\projQ=1-\projP$ is the projector orthogonal to the angle-resolved
density fluctuations, and $\smol^\dagger_\text{irr}$ is the irreducible
Smoluchowski operator (see Refs.~\cite{Liluashvili.2017,taggedpaper} for
details).

\subsection{Low-Density Solution}

We first recapitulate the solution at vanishing host-system density $n$.
Since Eqs.~\eqref{eq:mz} and \eqref{eq:mzs} have been
written such that the corresponding memory kernels are irreducible,
these memory kernels are at least of $\mathcal O(n)$, and can be dropped
in the dilute limit.
The low-density theory is thus given by
\begin{equation}\label{eq:lowdens}
  \partial_t\tilde{\bs\Phi}(q,t)+\tilde{\bs\omega}(q)\cdot
  \bs S^{-1}(q)\cdot\tilde{\bs\Phi}(q,t)=\bs0\,,
\end{equation}
where in leading order in the density, also $\bs S(q)=\bs 1$.
Formally this is solved by $\tilde{\bs\Phi}(q,t)=\exp[\tilde{\bs\omega}(q)\cdot\bs S(q)^{-1}t]\bs S(q)$, and the corresponding expression
holds for $\tilde{\bs\phi}(q,t)$.
We note in passing that an alternative representation of the exact
low-density solution is in terms of suitable eigenfunctions of the
Smoluchowski operator for a free \gls{ABP} \cite{Kurzthaler.2016,Kurzthaler.2018},
the Mathieu functions in \gls{2D}.
It can be shown that the two representations are indeed equivalent \cite{taggedpaper}.

Specializing the tagged-particle equivalent of Eq.~\eqref{eq:lowdens}
to the positional density correlator $\phi^s_{00}(q,t)$,
making use of the tri-diagonal structure of $\tilde{\bs\omega}^s(q)$,
we get
\begin{equation}\label{eq:phi00ld}
  \partial_t\phi^s_{00}(q,t)+q^2D_t\phi^s_{00}(q,t)
  =\sum_\pm\frac{iqv_0^s}2\tilde\phi^s_{\pm1,0}(q,t)\,.
\end{equation}
The low-density limit of the dipole correlator $\tilde\phi^s_{\pm1,0}(q,t)$
is similarly given by
\begin{multline}\label{eq:phi10ld}
  \partial_t\tilde\phi^s_{\pm1,0}(q,t)+\tilde\omega^s_{\pm1,\pm1}(q)
  \tilde\phi^s_{\pm1,0}(q,t)
  +\tilde\omega^s_{\pm1,0}(q)\tilde\phi^s_{00}(q,t)
\\
  +\tilde\omega^s_{\pm1,\pm2}(q)\tilde\phi^s_{\pm2,0}(q,t)=0\,.
\end{multline}
In the low-$q$ limit, observe that
\begin{equation}
  \hat\phi^s_{\pm10}(t)=\lim_{q\to0}\frac1q\tilde\phi^s_{\pm10}(q,t)
\end{equation}
is well-defined and nontrivial
since the $(ll')=(\pm1,0)$ correlator is of $\mathcal O(q)$
by the rotation property Eq.~\eqref{eq:phitransform}.
In leading order for $q\to0$, we can replace $\tilde\phi^s_{00}(q,t)=1$ and
drop the last term in Eq.~\eqref{eq:phi10ld}, because $\tilde\phi^s_{\pm2,0}(q,t)
=\mathcal O(q^2)$ due to the rotation property. Inserting the explicit
expressions of $\tilde{\bs\omega}^s(q)$, one obtains finally the coupled
differential equations that determine the \gls{MSD},
\begin{subequations}\label{eq:msd}
\begin{gather}
  \partial_t\delta r^2(t)=4D_t^s-2\sum_\pm(iv_0^s)\hat\phi^s_{\pm1,0}(t)\,,\\
  \partial_t\hat\phi^s_{\pm1,0}(t)+D_r^s\hat\phi^s_{\pm1,0}(t)
  =\frac{iv_0^s}2\,.
\end{gather}
\end{subequations}

Equations \eqref{eq:msd} are readily solved. From
\begin{equation}
  \hat\phi^s_{\pm1,0}(t)=\frac{iv_0^s}{2D_r^s}\left(1-e^{-D_r^st}\right)
\end{equation}
one gets finally
\begin{equation}\label{eq:msdfree}
  \delta r^2(t)=4D_t^st\left(1+\Pe^s\left(1+\frac{e^{-D_r^st}-1}{D_r^st}\right)
  \right)\,,
\end{equation}
where we have introduced the P\'eclet number $\Pe^s={v_0^s}^2/2D_r^sD_t^s$.
This indeed agrees with the well-known solution for the mean-squared
displacement of a single \gls{ABP} obtained by direct solution of the
corresponding stochastic differential equation \cite{Howse.2007}.
It should be noted that for this re-derivation of the result based on the
Mori-Zwanzig equations, it is essential that the projector $\projP$ includes
all angular-index values $l$, even if the \gls{MSD} refers to the
$l=l'=0$ motion only. If one were to project only onto the $l=0$ density
fluctuations, the memory kernel would not vanish in the low-density limit.

It is worth recalling the typical time- and length-scales that are
inherent in Eq.~\eqref{eq:msdfree}. After a short-time passive-Brownian
regime, $\delta r^2\simeq 4D_t^st$ for $t\ll\tau_v$, a ballistic
regime, $\delta r^2\simeq (v_0^s)^2t^2$ for $\tau_v\ll t\ll\tau_l$ is followed
by a final cross-over to an activity-enhanced diffusive regime,
$\delta r^2\simeq 4D_t^s(1+\Pe^s)t$ for $t\gg\tau_l$.
Here, $\tau_v$ and $\tau_l$ mark the characteristic time scales of the
free \gls{ABP}, and they are associated with length scales
$\ell_{v,l}=\sqrt{\delta r^2(\tau_{v,l})/4}$. From balancing the asymptotic
expressions against each other, one obtains (dropping $s$ superscripts for
convenience)
\begin{subequations}\label{eq:msdlengths}
\begin{align}
  \tau_v&=\frac{2}{D_r\Pe}\,, & \ell_v&=\frac{2D_t}{v_0}\,;\\
  \tau_l&=\tau_v(1+\Pe)\,,    & \ell_l&=\ell_v+\frac{v_0}{D_r}\,.
\end{align}
\end{subequations}

\subsection{General equations for the \gls{MSD}}

To derive expressions for the \gls{MSD}, Eq.~\eqref{eq:mzs} needs
to be evaluated in the limit $q\to0$. This requires an expression for
the inverse of the frequency matrix $\bs\omega_T^s(\vec q)$
that appears in the definition of the memory kernel. The simple
tri-diagonal structure,
\begin{equation}\label{eq:omegaTs}
  \tilde\omega_{T,ll'}^s(\vec q)=\delta_{ll'}q^2D_t^s
  -\delta_{|l-l'|,1}\frac{iqv_0^s}{2}\,,
\end{equation}
allows to derive an analytical expression \cite{taggedpaper},
\begin{equation}\label{eq:omegaTsinv}
  (\tilde\omega_T^s)^{-1}_{ll'}=\frac1{\sqrt{(D_t^sq^2)^2+(v_0^sq)^2}}
  \left(\frac{iqv_0^s}{D_t^sq^2+\sqrt{(D_t^sq^2)^2+(v_0^sq)^2}}\right)^{|l-l'|}\,.
\end{equation}
The result is readily verified by direct multiplication.
It demonstrates an intricate feature of the low-$q$ limit of the theory:
the $q\to0$ asymptotes of Eq.~\eqref{eq:omegaTsinv} are distinct for
the passive case, $v_0^s=0$, and the active case, $v_0^s\neq0$. Explicitly,
one gets
\begin{align}
  (\tilde\omega_T^s)^{-1}_{ll'}&\simeq\frac1{D_t^sq^2}\delta_{ll'}\,,&&\text{passive,}\\
  (\tilde\omega_T^s)^{-1}_{ll'}&\simeq\frac1{v_0^sq}
  \left(1-|l-l'|\frac{q}{q_*}\right)+\mathcal O(q)\,,&&\text{active,}
\end{align}
with $q_*=v_0^s/D_t^s$.
This peculiar feature encodes that even an arbitrarily small activity
of the tracer is felt, given that one probes length scales that are sufficiently
large.
In order to establish the correct $q\to0$ asymptote, the inversion of the
matrix $\tilde{\bs\omega}_T^s(q)$ needs to be performed on the
infinite-dimensional algebra of matrices labeled by angular-mode indices.
Care has to be taken when introducing a cutoff for the angular modes,
as is customary in numerical implementations: the cutoff matrix
$\tilde{\bs\omega}_T^s(q)$ has an inverse that has the wrong $q\to0$
asymptote (either $\sim1/q^2$ or constant for the $(00)$ element,
depending on whether the
cutoff is performed at even or odd angular modes). The recognition that
the inverse has to be performed \emph{before} introducing a cutoff
is crucial in deriving the correct \gls{MSD} equations of motion.

To complete the derivation, the $q\to0$ limits of the memory kernels
$\tilde{\bs M}^s(q,t)$ are needed.
Note first the case of a passive tracer: there, $\tilde{\bs\phi}^s(q,t)$
and $\tilde{\bs\omega}^s(q)$ remain diagonal matrices,
and $\tilde M^s_{00}(q,t)=\mathcal O(q^2)$. Hence, only the memory integral
involing $\tilde m^s_{00}$ remains in the equation determining the \gls{MSD}.
For an active tracer, one needs to recognize that for $l>0$,
$\tilde\phi^s_{l0}(q,t)=\mathcal O((iqv_0^s)^{|l|})$,
and that the memory kernels $\tilde M^s(q,t)$ are of at least
$\mathcal O(q^0)$. Combined with the terms $1/q$ and $1/q^2$ that
appear in $\tilde{\bs\omega}_T^{-1}(q)$, the potentially relevant
terms are $\tilde m^s_{ll'}(t)$ for $l'=\pm1$. In the $q\to0$ limit
the powers of $q$ stemming from $\tilde\phi^s_{l0}(q,t)$ leave
the quantities $\hat m_{ll'}(t)=\lim_{q\to0}\tilde m_{ll'}(q,t)/q^{|l-l'|}$
for $|l-l'|\le1$ as the relevant memory kernels \cite{taggedpaper}.

Hence, we arrive at the coupled integro-differential equations that
describe the time-evolution of the \gls{ABP}-\gls{MSD} in a crowded
environment,
\begin{subequations}\label{eq:msdmct}
\begin{multline}\label{eq:msdmct0}
  \partial_t\delta r^2(t)
  +\int_0^tdt'\,\hat m^s_{00}(t-t')\delta r^2(t')
  =4D_t^s\\-2\sum_\pm(iv_0^s)\hat\phi^s_{\pm1,0}(t)
  +4\sum_\pm\int_0^tdt'\,\hat m^s_{0,\pm1}(t-t')(\partial_{t'}+D_r^s)\hat\phi^s_{\pm1,0}(t)\,,
\end{multline}
together with
\begin{multline}
  \partial_t\hat\phi^s_{\pm1,0}(t)+D_r^s\hat\phi^s_{\pm1,0}(t)
  =\frac{iv_0^s}{2}\\
  -2\int_0^tdt'\hat m^s_{\pm1,\pm1}(t-t')(\partial_{t'}+D_r^s)\hat\phi^s_{\pm1,0}(t')\,.
\end{multline}
\end{subequations}
Equations \eqref{eq:msdmct} are the generalizations of the low-density
result, Eqs.~\eqref{eq:msd}, to arbitrary density of the host system.

Further evaluation requires specific expressions for the memory kernels.
The results that we discuss in the following are obtained by
employing \gls{ABPMCT}. This theory approximates the memory kernel
$\tilde{\bs M}^s(q,t)$ as a bilinear functional involving the density
correlation functions,
\begin{equation}\label{eq:mctMs}
  M^s_{l_1l_1'}(\vec q,t)\approx\delta_{\vec q\vec q_1}\delta_{\vec q\vec q_{1'}}
  \sum_{l_3l_4\vec q_3\vec q_4}\mathcal W^s_{l_1l_{1'}l_3l_4}(\vec q_1,\vec q_3)
  \Phi_{l_3,0}(\vec q_3,t)\phi^s_{l_4l_2}(\vec q_4,t)\,.
\end{equation}
Equations~\eqref{eq:mz} and \eqref{eq:mzs} are then closed by the
\gls{ABPMCT} expression for the collective memory kernel,
\begin{equation}\label{eq:mctM}
  M_{l_1l_{1'}}(\vec q,t)\approx\delta_{\vec q\vec q_1}\delta_{\vec q\vec q_{1'}}
  \sum_{343'4'}\mathcal V^\dagger_{134}\Phi_{33'}(t)\Phi_{44'}(t)
  \mathcal V_{1'3'4'}\,.
\end{equation}
The vertices $\mathcal V$, $\mathcal V^\dagger$, and $\mathcal W$ are
given in terms of the equilibrium static structure factor $S(q)$ of the
passive system; the \gls{ABPMCT} expressions can be found in
Ref.~\cite{taggedpaper}.
As also shown there, a somewhat tedious procedure confirms the
$q\to0$ limit of the evolution equations, Eq.~\eqref{eq:msdmct},
within \gls{ABPMCT}.

The first line of Eq.~\eqref{eq:msdmct0} corresponds to the expression
derived earlier for the \gls{MSD} of a passive tracer in a dense
system \cite{Fuchs.1998,Bayer.2007}. In a passive host system, the memory kernel
$\hat m_{00}^s(t)$ is a completely monotone function, i.e., it is
positive and a continuous superposition of purely relaxing exponentials;
as a consequence, the \gls{MSD} of a passive tracer in a passive host system
is always slowed down compared to free diffusion, and it follows an
increasingly pronounced sub-diffusive regime with increasing host-system
density.
Within \gls{ABPMCT}, the effect of activity in the host system enters
through a modification of the dynamics in the collective density
correlation function $\tilde\Phi_{00}(p,t)$, cf.\ Eq.~\eqref{eq:mctMs}.
Since within the theory, the collective density correlations decay faster
with increasing activity \cite{Liluashvili.2017,taggedpaper},
this suggests enhanced diffusivity for the tracer particle due to active
host-particle motion.
However, there is also an explicit dependence of the coupling
vertices $\mathcal W$ on the host-system activity, whose structure
admits more complex solutions; in particular, as we will discuss below,
there appears a super-diffusive regime in the \gls{MSD} even of a passive
tracer particle.
This is remarkable since it demonstrates the non-equilibrium nature
of the motion: For dynamics driven by the equilibrium Smolchowski operator,
it can be shown exactly, that there can be no superdiffusive regimes
in the \gls{MSD}. A brief proof for this statement is given in
Appendix~\ref{sec:eqmsd}.

\subsection{Numerics and Simulation}

In the following, we discuss exemplary features of the solutions of
Eqs.~\eqref{eq:msdmct} for the case of \gls{2D} hard disks of diameter
$\sigma$. The single control parameter for the passive system is then
the dimensionless number density $n$, or the packing fraction
$\varphi=(\pi/4)n\sigma^2$.
The size $\sigma$ sets the unit of length, and $\sigma^2/D_t$ the unit
of time. We consider tracer particles that are of the same size as the host
particles, and unless otherwise noted, $D_t=D_t^s=1$ and $D_r=D_r^s=1$
are chosen.
We briefly outline the numerical evaluation of Eqs.~\eqref{eq:msdmct}
within \gls{ABPMCT}, and the \gls{EDBD} simulation scheme whose results
we discuss in the following. Further details can be found in
Refs.~\cite{taggedpaper,ReichertPhD}.

The \gls{ABPMCT} expressions for the memory kernels were discretized on
an equidistant grid of $128$ points in wave numbers $q$, with cutoff
$40\sigma$. A cutoff of $L=1$ was used for the angular-mode indices,
which allows to study the regime of not too large self-propulsion velocities
in the theory. The current implementation suffers from numerical instabilities
at large $v_0^s$ that arise from the specific details of the implementation
of the integral solver, and the prohibitive memory and runtime requirements
for matrices with larger cutoff $L$. We thus restrict the discussion of the
theory to $v_0^s\le8\,D_t/\sigma$.
To obtain the dynamics, \gls{ABPMCT} requires as input quantity the
equilibrium static structure factor of the system, for which we use a
recent result from \gls{DFT} \cite{Thorneywork.2018}.

Simulations were carried out with $N=625$ particles with uniform size
polydispersity to avoid crystallization (standard deviation $0.2\sigma$). The
\gls{EDBD} scheme is essentially a rejection-free Monte Carlo approach
\cite{Scala.2007} where random Gaussian displacements are chosen at
every time step in order to implement Brownian motion, and potential
particle overlaps are resolved by performing elastic collisions between
the particles. The inclusion of a suitable drift in the Gaussian displacements
implements the active motion \cite{Ni.2013}.
Simulation trajectories were equilibrated for at lest $10^4$ time steps,
and averaged over at least $200$ realizations, and over initial times
in the stationary regime.

\section{Results}

\subsection{Passive Tracer in Passive Bath}

\begin{figure}
\includegraphics[width=\linewidth]{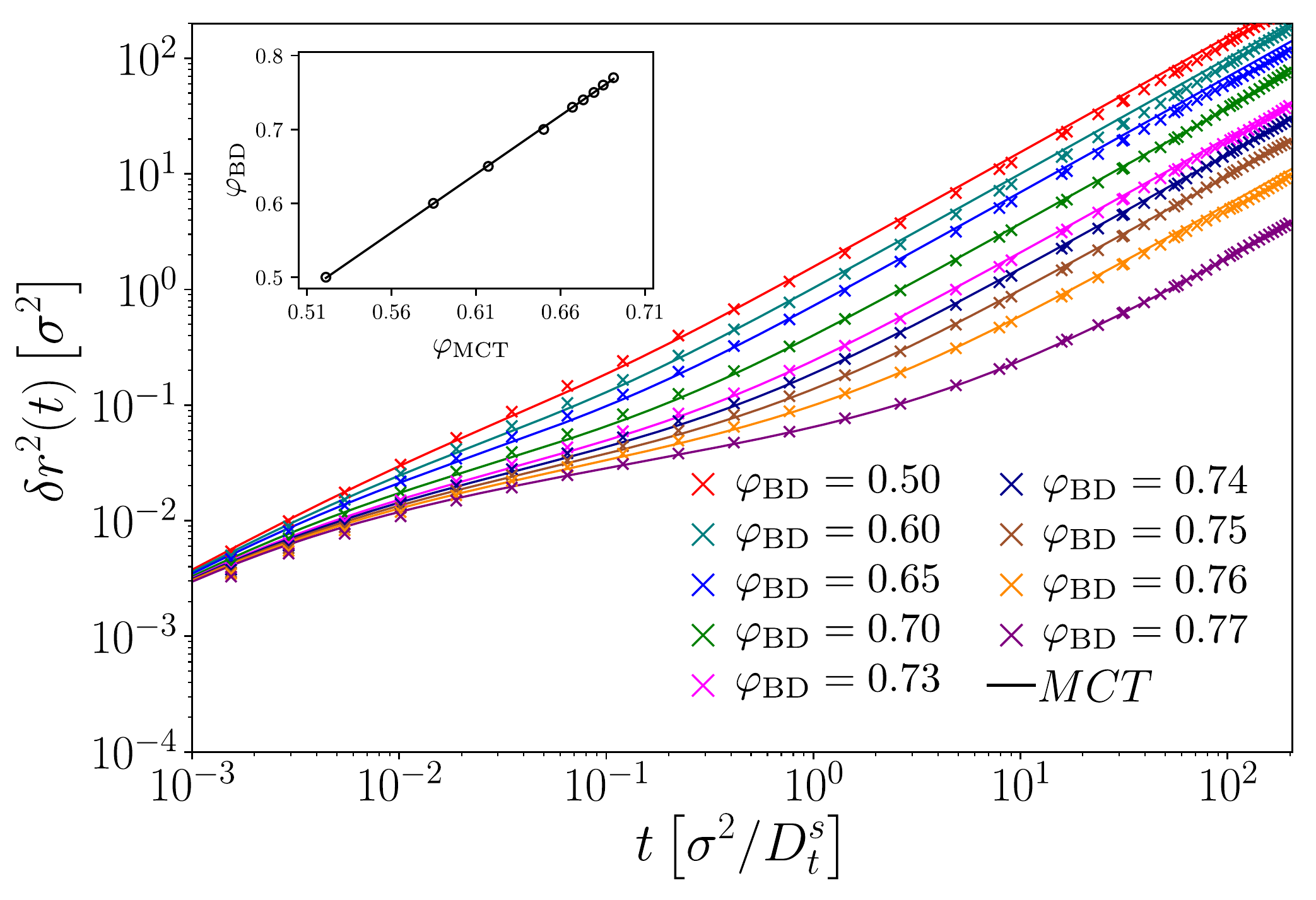}
\caption{\label{fig:passive}
  Mean-squared displacements $\delta r^2(t)$
  of a passive hard disk in a bath of passive
  hard disks, at various packing fractions $\varphi_\text{BD}$ as indicated.
  Symbols are results from Brownian-dynamics computer simulations,
  lines are predictions of mode-coupling theory with packing fraction
  $\varphi_\text{MCT}$
  adjusted to account for the different glass-transition points.
  The inset shows the employed relation between $\phi_\text{MCT}$ and
  $\phi_\text{BD}$, which follows a linear variation.
}
\end{figure}

To establish a baseline for the comparison of \gls{MCT} with our simulation
data, we first briefly demonstrate the results obtained for a passive
tracer particle in a passive host system (Fig.~\ref{fig:passive}).
In this comparison, we follow a well-established procedure to account for the
fact that the theory predicts dynamical arrest at a packing fraction
$\phi_c$ whose numerical value is different from the one seen in simulation.
In particular, with the specific choice of numerical parameters and
\gls{DFT} static structure factor in our \gls{MCT} calculations, we obtain
$\phi_{c,\text{MCT}}\approx0.699$ in close agreement with
the result obtained from a \gls{MHNC} approximation of $S(q)$ and
a somewhat different numerical integration scheme for the memory
kernel \cite{Bayer.2007}.
From the \gls{EDBD} simulations we estimate
$\phi_{c,\text{BD}}\approx0.78$.

Since the relevant parameter describing the long-time dynamics in \gls{MCT}
is asymptotically linearly related to the control-parameter distance
$\varepsilon=\varphi-\varphi_c$, up to a prefactor or $\mathcal O(1)$,
one expects that the theory describes the dynamics of the system after
mapping the packing fraction $\varphi_\text{BD}$ asymptotically
linearly to a (smaller) packing
fraction $\varphi_\text{MCT}$ that enters the \gls{MCT} calculations.
This has been discussed in detail for \gls{3D} hard spheres \cite{Weysser.2010}.
Note that the mapping of packing fractions that we use here differs somewhat
from the one that gives best agreement between theory and simulation for
the density correlation functions at finite $q$ \cite{taggedpaper}; this also
has previously been discussed in \gls{3D} system \cite{Weysser.2010} and is attributed to
a further quantitative error of the \gls{MCT} approximation in the low-$q$
regime of the memory kernel.

Keeping this caveat in mind, we find (cf.\ Fig.~\ref{fig:passive})
that after the adjustment of packing fractions, \gls{MCT} provides an
excellent quantitatively accurate description of the \gls{MSD} obtained
from our \gls{EDBD} simulations in the regime of packing fractions approaching
$\varphi_c$.

The \gls{MSD} show the typical features known from glass-forming Brownian
systems: after a short-time diffusive asymptote, $\delta r^2\simeq 4D_tt$
(in our systems where hydrodynamic interactions are absent),
a regime of subdiffusive
motion, i.e., of sublinear growth in $\delta r^2(t)$ as a function of time,
marks the transient cageing of particles by their neighbors. On the fluid
side of the glass transition that we study here, $\varphi<\varphi_c$, the \gls{MSD}
eventually crosses over to a long-time diffusive asymptote,
$\delta r^2\simeq 4D_t^Lt$, where $D_t^L(\varphi)$ is the long-time translational
self-diffusion coefficient that decreases strongly as $\varphi$ is increased
and is predicted by \gls{MCT} to vanish at $\varphi_c$.
Hence, as the
density of the system is increased, the cageing regime extends to increasingly
long times, and in the ideal glass predicted by \gls{MCT}
the \gls{MSD} arrests to a finite plateau value,
$\delta r^2(t)\simeq 4\ell_c^2$ as $t\to\infty$ for $\varphi\ge\varphi_c$.
The length scale $\ell_c$ quantifies a typical ``cage size'' in the glass,
and by a simple argument due to Lindemann is expected to be some fraction
of the particle size, typically around $10\%$.
Indeed, from inspection of Fig.~\ref{fig:passive} we estimate
$\ell_c\approx0.087\sigma$ in our system, in excellent agreement with the
\gls{MCT} prediction.

We restrict the discussion in the following to densites $\varphi_\text{BD}\le0.77$, where the system still represents a fluid.
At larger densities, the simulated \gls{MSD} do not show kinetic arrest
in our \gls{2D} system, which is expected on the grounds of
Mermin-Wagner fluctuations that provide an additional relaxation channel
\cite{Illing.2017,Vivek.2017,Shiba.2016} that is not accounted for in
the theory.

\subsection{Active Tracer in Passive Bath}

\begin{figure}
\includegraphics[width=\linewidth]{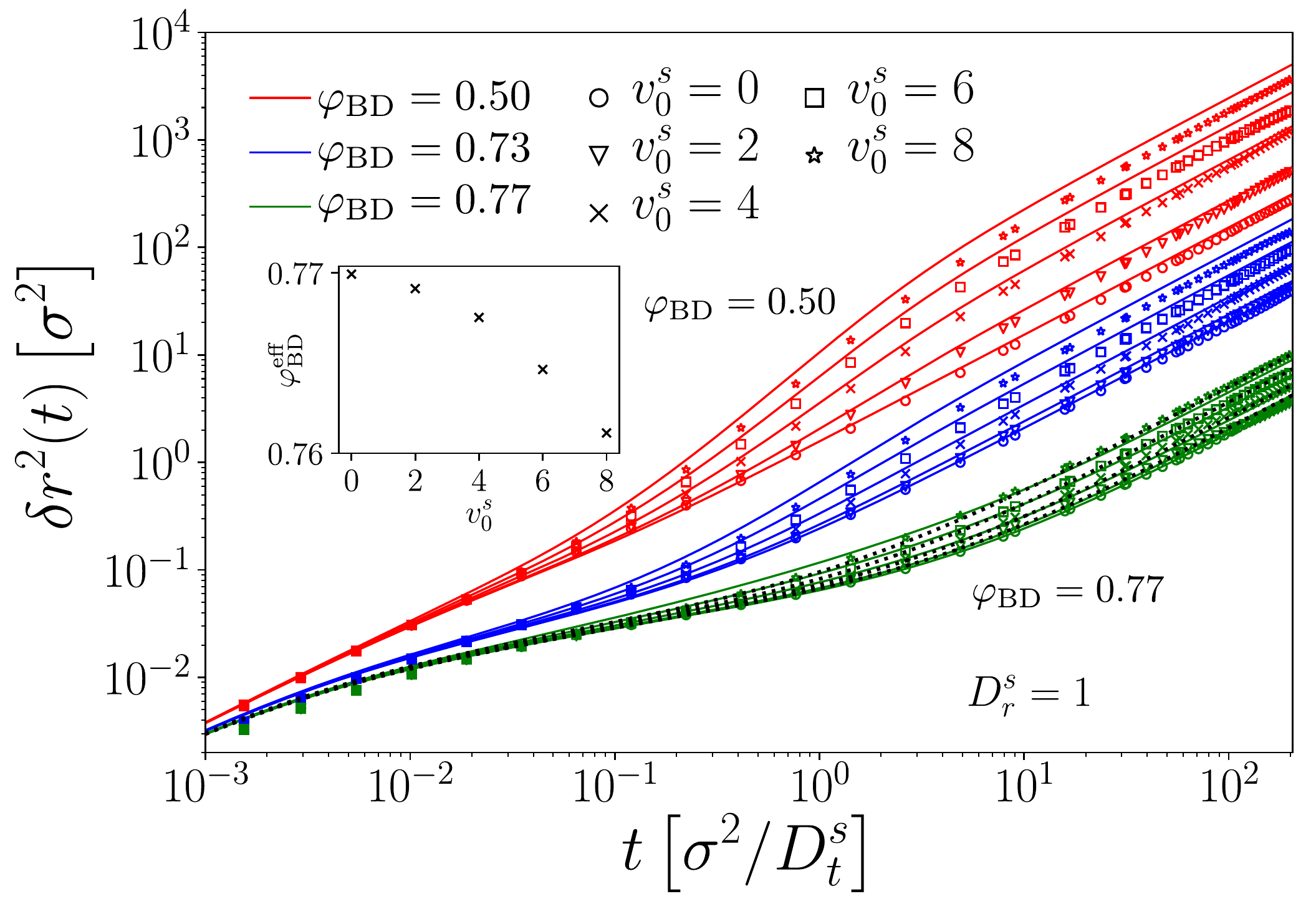}
\caption{\label{fig:active-passive}
  \Acrfullpl{MSD} $\delta r^2(t)$ of a single active Brownian
  particle in a bath of passive hard disks. Symbols are results from
  Brownian-dynamics simulations at packing fraction $\varphi_\text{BD}$
  as labeled (color-coded), for different self-propulsion velocities
  $v_0^s$ of the tracer (as labeled, increasing from bottom to top in
  each group of curves). Lines are results from MCT, with the adjustment
  of packing fractions shown in Fig.~\protect\ref{fig:passive}.
  Dotted lines for $\varphi=0.77$ show fits using the \gls{MSD}
  of a passive tracer at a reduced host-packing fraction
  $\varphi_\text{eff}(v_0^s)$ (inset).
}
\end{figure}

Having established the accuracy of \gls{MCT} for the passive system after
a suitable mapping of densities, we now turn to the dynamics of a single
active tracer in a system of passive hard disks.
The \gls{EDBD} simulation results and the predictions of \gls{MCT} without
further adjustment of parameters again are in very good agreement
(Fig.~\ref{fig:active-passive}), in the parameter range of $v_0^s$ within
which we can obtain numerically stable solutions of the \gls{MCT} equations
of motion.

We exemplarily discuss the case $v_0^s=8$ for the densities $\varphi=0.50$
and $\varphi=0.77$. Recall from Eq.~\eqref{eq:msdlengths} that for a
free \gls{ABP} with $v_0^s=8$, a ballistic regime appears in the \gls{MSD}
for $4\ell_v^2=1/4\ll\delta r^2\ll1089/4=4\ell_l^2$. In the case of a moderately
dense host system, $\varphi=0.50$, the \gls{MSD} or the \gls{ABP} tracer
still evidences this regime of persistent active motion. Yet, as the
density of the passive host system is increased to $\varphi=0.77$, both
our simulations and theory indicate that a superdiffusive regime is not
encountered any more. Here, the strong influence of the cage effect
that is responsible for the glassy dynamics suppresses the persistent
motion of the single active tracer particle.

Recall that $\ell_c\approx0.087\ll\ell_v$ for the choice of parameters that
we discuss here. It is thus plausible that the passive-host dynamics
suppresses the persistent active motion of the tracer at sufficiently
high densities, and as a result, the \gls{MSD} at $\varphi=0.77$ qualitatively
appear as one would also observe for a passive tracer particle; the
activity in this regime is only seen through an enhanced long-time
diffusion.
In fact, the data for $\varphi=0.77$ can be fit with the \gls{MSD}
of a passive tracer, at an effectively reduced host-system packing fraction
$\varphi^\text{eff}(v_0^s)$ (dotted lines in Fig.~\ref{fig:active-passive}).
The $\varphi^\text{eff}$-vs-$v_0^s$ relation (inset of the figure)
shows the expected quadratic dependence on $v_0^s$ that conforms to
the $v_0^s\mapsto-v_0^s$ symmetry of the ensemble. In this system
of active hard disks, where temperature is irrelevant (and only sets an
overall time scale of the motion), $\varphi^\text{eff}$ can be seen as the
analog of an effective temperature, $T_\text{eff}(v_0^s)-T_0\propto(v_0^s)^2$
in the sense that active motion reduces the coupling strength to the bath.
From the low-density solution, Eq.~\eqref{eq:msdfree}, one would
identify $T_\text{eff}-T_0=\Pe$, and the corresponding enhancement
of diffusivity for the parameters exemplified in Fig.~\ref{fig:active-passive}
is a factor of $32$. For the passive long-time dynamics, a change in
(effective) temperature would bring about an even larger change in the
long-time diffusivity, while the enhancement seen in Fig.~\ref{fig:active-passive}
for $\varphi=0.77$ is only around a factor $2$. This clearly indicates
the limitations of the effective-temperature picture \cite{Berthier.2013}
for an \gls{ABP} in a crowded medium.

\begin{figure}
\includegraphics[width=\linewidth]{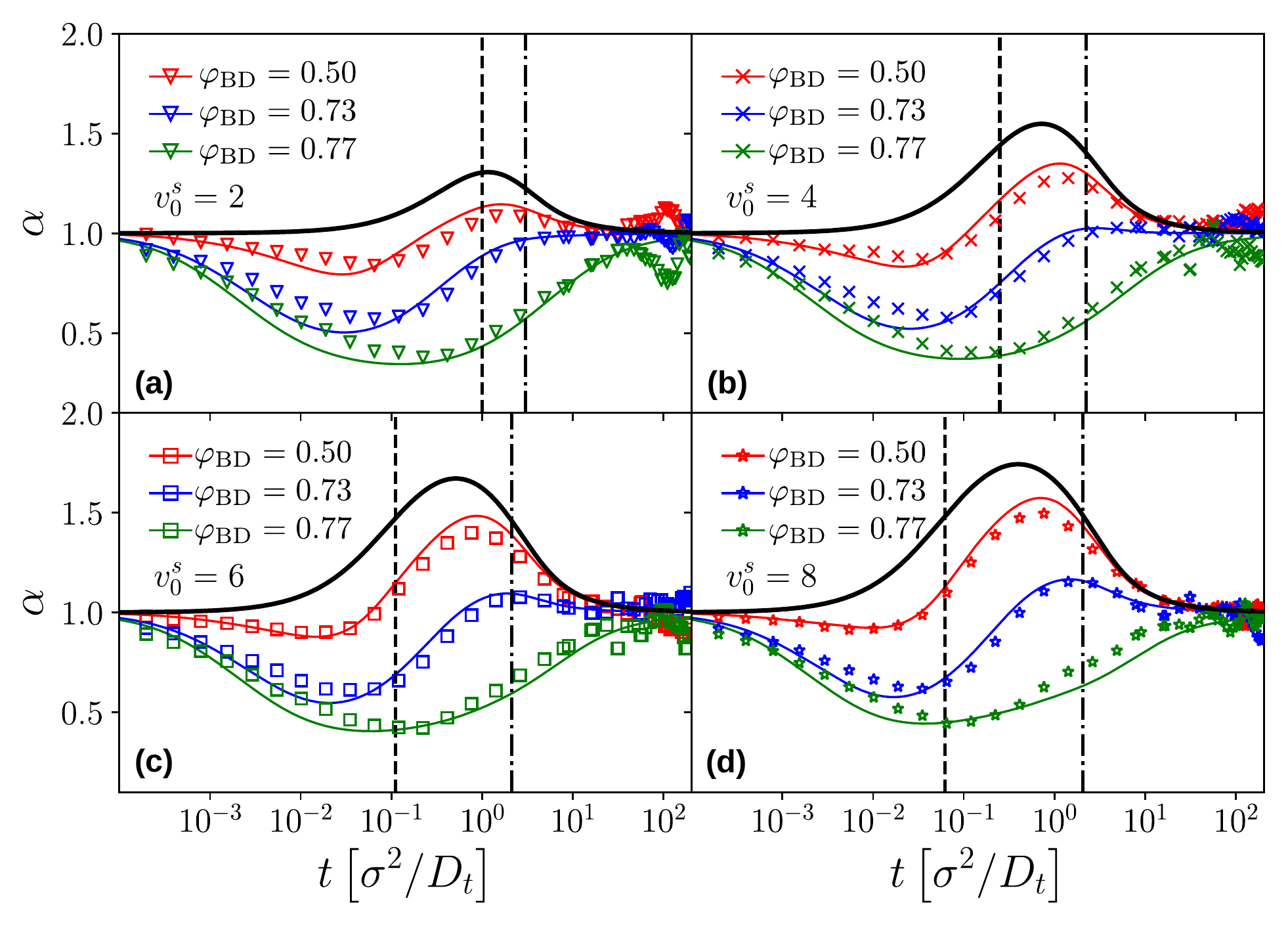}
\caption{\label{fig:active-passive-exp}
  Effective epxonents $\alpha(t)=d\log\delta r^2(t)/d\log t$ obtained
  from the MSD of an active tracer in a passive host system,
  corresponding to the data shown in Fig.~\protect\ref{fig:active-passive}.
  Panels (a)--(d) show the results for the different self-propulsion
  velocities $v_0^s$ of the tracer. Symbols are BD simulation results,
  lines are MCT results. Thick black lines correspond to the analytical
  solution for a free active particle. Vertical dashed and dot-dashed
  lines indicate the time scales $\tau_v$ and $\tau_l$ characterizing the
  free-particle MSD, cf.\ Eq.~\protect\eqref{eq:msdlengths}.
}
\end{figure}

The appearance and disappearance of superdiffusive motion is best seen
by the effective power-law exponents associated to the \gls{MSD}.
Recall that for any function that is a power law, its logarithmic
derivative $\alpha(t)=d\log\delta r^2(t)/d\log t$ will be constant and equal
to the power-law exponent. The effective exponents
$\alpha(t)$ corresponding to the \gls{MSD} shown in Fig.~\ref{fig:active-passive}
confirm the interplay between subdiffusive cage motion, $\alpha<1$, and
super-diffusive persistent active motion, $\alpha>1$, for the active
tracer in the dense passive host system (Fig.~\ref{fig:active-passive-exp}).
For the choice of parameters represented in the figure,
the cage effect sets in at times earlier than the free-particle crossover
to persistent motion, since $\ell_c\ll\ell_v$. As a result, the effective
exponent in all cases follows an S-shaped curve that first drops to
values below unity, and increases to values above unity in the time window
$\tau_v\ll t\ll\tau_l$ that follows the cageing dynamics. The cross-over
where the subdiffusive cage motion is modified by the tracer activity
is, even at the density $\varphi=0.73$, well predicted by $\tau_v$ (vertical
dashed lines in Fig.~\ref{fig:active-passive-exp}).

\begin{figure}
\includegraphics[width=\linewidth]{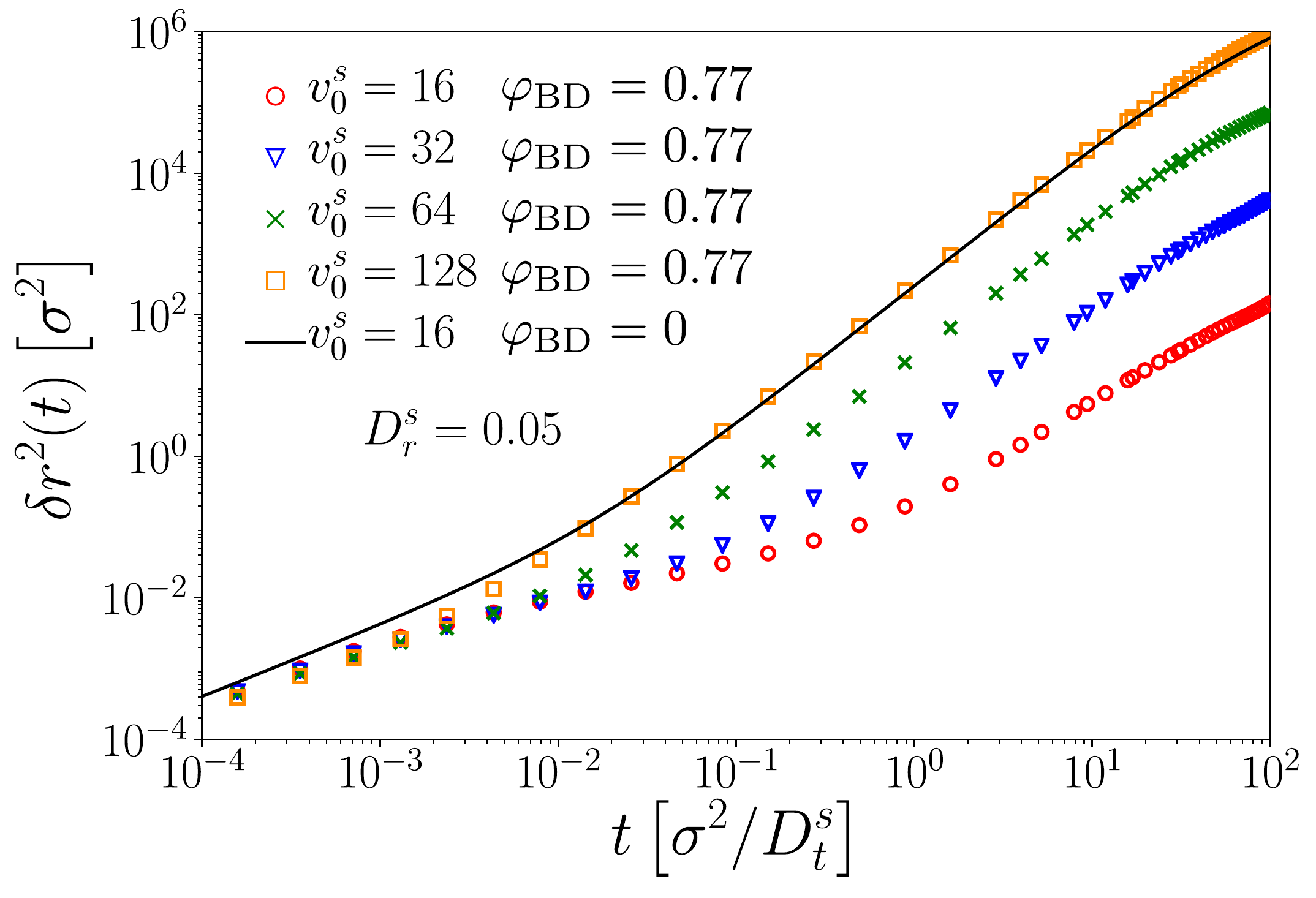}
\caption{\label{fig:active-passive-large}
  Mean-squared displacements $\delta r^2(t)$ of a single active Brownian
  particle in a bath of passive hard disks of packing fraction
  $\varphi=0.77$, for self-propulsion velocities $v_0^s$ as labeled,
  and for $D_r^s=0.05$.
  Symbols are BD simulation results. A solid line indicates the
  MSD of a free active Brownian particle with a self-propulsion velocity
  of $v_{0,\text{free}}^s=16$.
}
\end{figure}

At very large $v_0^s$, one expects the transition to persistent motion
(on time scale $\tau_v$, respectively length scale $\ell_v$) to occur
even before cageing becomes effective. Currently, the required large
$v_0^s$ do not allow us to solve the \gls{MCT} equations reliably. We thus
turn to \gls{EDBD} simulations in this regime (Fig.~\ref{fig:active-passive-large}).
Indeed, even at the density $\varphi=0.77$ for which the passive host system
induces subdiffusive cageing motion over about three decades in time for
the passive or moderately active tracers, we observe in our \gls{EDBD}
simulations for large $v_0^s$ an increasingly rapid cross-over to
superdiffusive motion that replaces the subdiffusive regime entirely
once $\ell_v\ll\ell_c$. This is exemplified for $v_0^s\gtrsim32$ by
the \gls{EDBD} data (Fig.~\ref{fig:active-passive-large}; where we have also
set $D_r^s=0.05\,D_t/\sigma^2$ to emphasize the effect). In essence, strong activity
of sufficiently large persistence length eliminates the cage effect for
the active tracer. In the simulations it appears that as $v_0^s$ is further
increased, one essentially observes the motion of a free \gls{ABP},
with a density-renormalized swim speed.
To exemplify this, we compare the \gls{EDBD} results for $v_0^s=128$ with
the free-particle \gls{MSD} for $v_{0,\text{eff}}^s=16$; both curves
agree closely (Fig.~\ref{fig:active-passive-large}).

It would be worth further investigation whether the active tracer undergoes
a delocalization transition even in the passive glass. For a tracer
that is driven by an external force of fixed direction, this effect
is known \cite{Puertas.2014} and has been studied in the framework
of \gls{MCT} \cite{Gazuz.2009,Gazuz.2013,Gruber1,Gruber2}. Here, the theory predicts that above a
certain threshold force, the tracer motion delocalizes (as indicated by
an \gls{MSD} that grows without bound even when the host system is glassy).
However, in the present theory the situation is less obvious, because
the active tracer always has a finite persistence time if $D_r>0$, and
the limit $D_r\to0$ does not necessarily commute with the long-time limit
of interest in studying glassy dynamics.

\begin{figure}
\includegraphics[width=\linewidth]{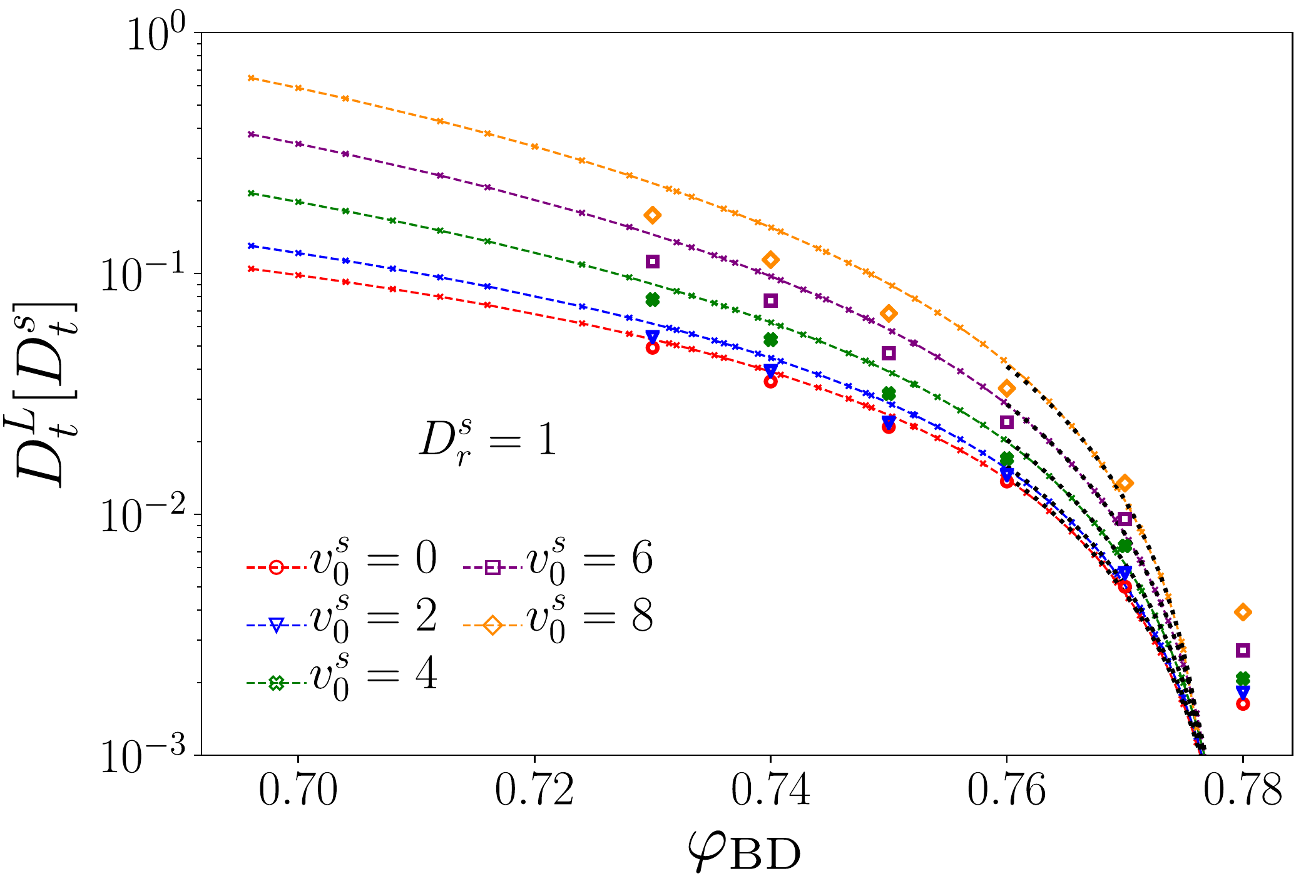}
\caption{\label{fig:active-passive-D}
  Long-time self-diffusion coefficient $D^L_t=\lim_{t\to\infty}\delta r^2(t)/4t$
  (in units of the tracer short-time diffusion coefficient $D_t^s$)
  obtained from the MSD of an active tracer in a passive hard-disk system,
  as a function of host packing fraction $\varphi$. Symbols are results from
  Brownian dynamics simulations for different tracer self-propulsion
  velocities $v_0^s$ as labeled. Small crosses connected by lines are
  MCT results with mapped packing fractions to match the dynamics of the
  passive system at $\varphi$ close to $\varphi_c$ (see text).
  The asymptotic power laws are indicated for $\varphi\ge0.76$
  as dotted lines.
}
\end{figure}

For the regime of moderate activity, the active tracer becomes,
within the theoretical idealization,
trapped in the passive host system at densities $\varphi\ge\varphi_c$. This
is expected because the cages possess a finite microscopic yield
strength \cite{Gazuz.2009}, and if the self-propulsion velocity $v_0^s$
translates into a force exerted by the active tracer that is below this
threshold (and also not infinitely persistent for finite $D_r$), cages
will not yield due to activity.

Approaching the glass transition, one thus expects the long-time motion
of the tracer to be quantified by a long-time diffusion coefficient
$D_t^L(\varphi)$ that approaches zero at $\varphi=\varphi_c$, and, by
\textit{bona fide}
extension of the well established asymptotic results of \gls{MCT} for
passive systems, vanishes as a power law close to the transition,
$D_t^L(\varphi)\sim|\varphi-\varphi_c|^\gamma$ for $\varphi\to\varphi_c$ from below.
The exponent $\gamma$ is a non-universal exponent emerging from the
asymptotic solution of the \gls{MCT} equations. In practice, the
\gls{MCT} description of the glass transition is an idealized one, and
one observes in simulations deviations from the power-law behavior
close to and above $\varphi_c$, rendering $D_t^L$ finite also there.

Our \gls{EDBD} simulations confirm this expectation (Fig.~\ref{fig:active-passive-D}). In the density window $0.76\lesssim\varphi\lesssim0.77$, the long-time
self-diffusion coefficient $D^L_t=\lim_{t\to\infty}\delta r^2(t)/4t$ follows
the power-law expected from \gls{MCT}. Deviations are seen for our simulations
at $\varphi=0.78$; as mentioned above, the appearance of long-range fluctuations
in the \gls{2D} system likely affects the data here, and we exclude this point
from our discussion.

As anticipated from Fig.~\ref{fig:active-passive}, increasing the self-propulsion velocity of the tracer enhances its long-time diffusion. The quantitative
agreement with \gls{MCT} deteriorates with increasing $v_0^s$, but the
qualitative behavior remains the same. With our choice of parameters,
discussing a change in $v_0^s$ at fixed $\varphi$ and fixed $D_r^s$, we observe
only a monotonic increase of $D^L_t$ with increasing $v_0^s$.
One should note that theory and simulations on a different model of
active particles, the \gls{AOUP}, demonstrate a non-monotonic variation
with activity \cite{Berthier.2019} that has also been reported from some experiments \cite{Klongvessa.2019a,Klongvessa.2019b}.

\begin{figure}
\includegraphics[width=\linewidth]{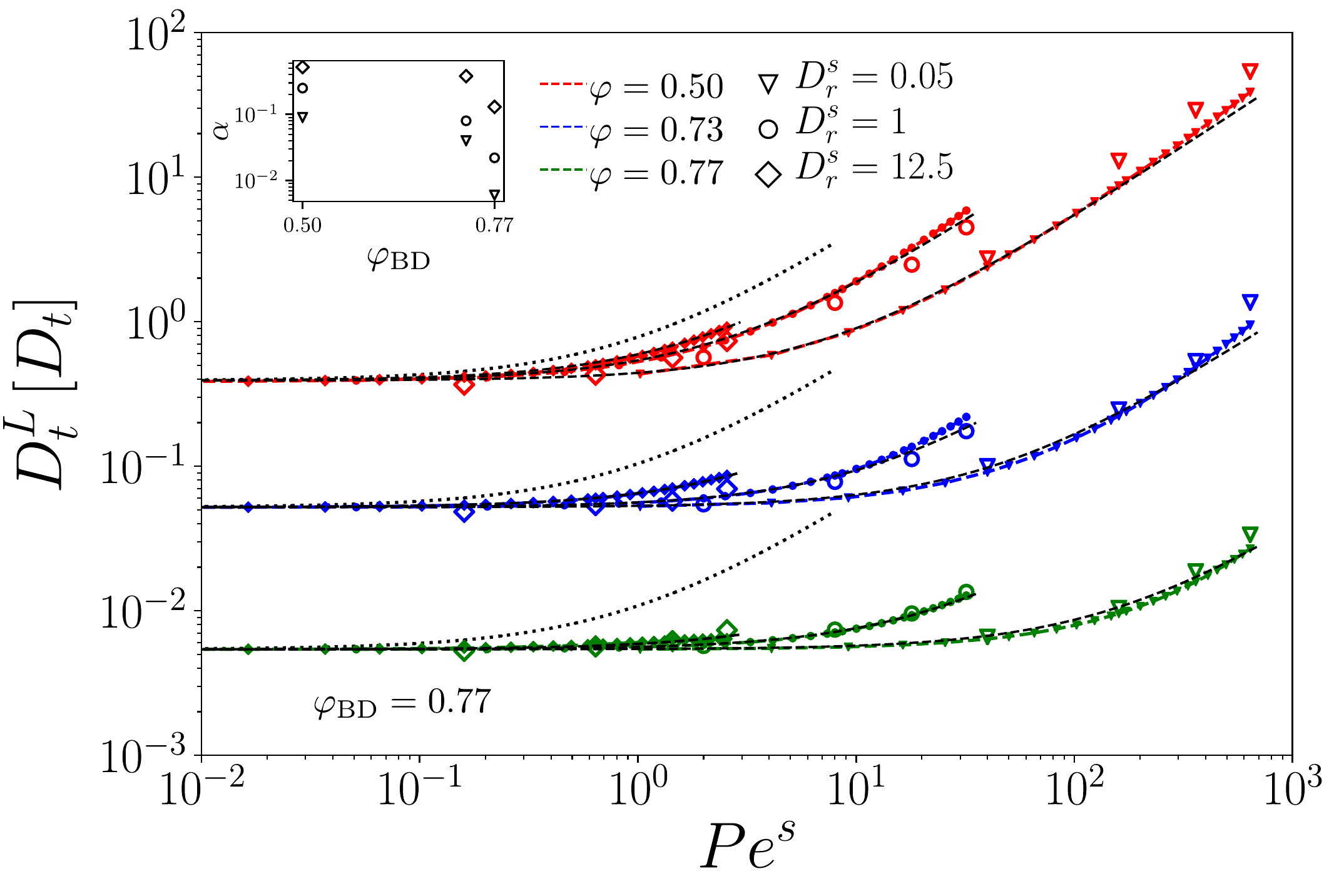}
\caption{\label{fig:active-passive-Dpe}
  Long-time diffusion coefficient $D_t^L(\varphi)$ of an active tracer in
  a passive hard-disk system, as a function of tracer P\'eclet number
  $\Pe^s=(v_0^s)^2/2D_r^sD_t^s$, for three different host-system packing
  fractions $\varphi$ as labeled. Open symbols are Brownian dynamics results,
  with different $D_r^s$ as labeled; small symbols connected with lines
  are MCT results. Dotted lines indicate the free-particle result
  $D_t^L=D_t^{L,0}(1+\Pe^s)$ scaled to the passive-tracer long-time diffusion
  coefficient $D_t^{L,0}$.
  Dashed lines are fits to the data with the empirical relation
  $D_t^L=D_t^{L,0}(1+\alpha\Pe^s)$, where $\alpha$ is shown in the inset.
}
\end{figure}

In the low-density regime, Eq.~\eqref{eq:msdfree} establishes that
in the long-time diffusive regime of the \gls{ABP}, activity only enters
through the dimensionless P\'eclet number $\Pe^s=(v_0^s)^2/2D_r^sD_t^s$.
In particular, one obtains $D_t^L=D_t^s(1+\Pe^s)$.
At high densities, this simple relation cannot be expected any more,
because the cage effect provides a further scale for the problem,
so that out of the two parameters that quantify the active motion
of the \gls{ABP}, $v_0^s$ and $D_r^s$, two independent dimensionless
numbers can be formed.

It is nevertheless instructive to check the scaling with $\Pe^s$.
Indeed, both theory and simulation demonstrate
that for any fixed $D_r^s$ the long-time diffusion coefficients are
of the form $\text{const.}+\Pe^s$ (symbols in Fig.~\ref{fig:active-passive-Dpe}).
This quadratic dependence on the self-propulsion velocity
is also expected from the symmetry of the ensemble under mapping
$v_0^s\mapsto-v_0^s$.
Yet, the prefactors depend on both $D_r^s$ and the packing fraction.
A simple-minded rescaled description that takes into account the reduced
diffusivity in the passive system,
$D_t^L=D_t^{L,0}(\phi)(1+\Pe^s)$ with $D_t^{L,0}(\phi)$ the density-renormalized
free diffusion of the passive particle, still fails (dotted lines
in Fig.~\ref{fig:active-passive-Dpe}).
An empirical rescaling, $D_t^L=D_t^{L,0}(\varphi)(1+\alpha(\varphi,D_r^s)\Pe^s)$
describes the data (dashed lines), and reveals two trends for the
rescaling factor $\alpha$: it decreases with increasing host-system
density, i.e., the enhancement of long-time diffusion at given
tracer-P\'eclet number becomes weaker. The prefactor $\alpha$ also increases
with increasing $D_r^s$ at fixed packing fraction.
This might indicate a limit of $D_r^s\to\infty$ and $v_0^s\to\infty$
at fixed $\Pe^s$, where the active tracer recovers effectively-free motion
with a renormalized Brownian diffusion coefficient due to the dense host
system, and it is compatible with the limit of the passive tracer
particle ($D_r^s\to0$ at fixed $\Pe^s$ which implies $v_0^s\to0$).

\begin{figure}
\includegraphics[width=\linewidth]{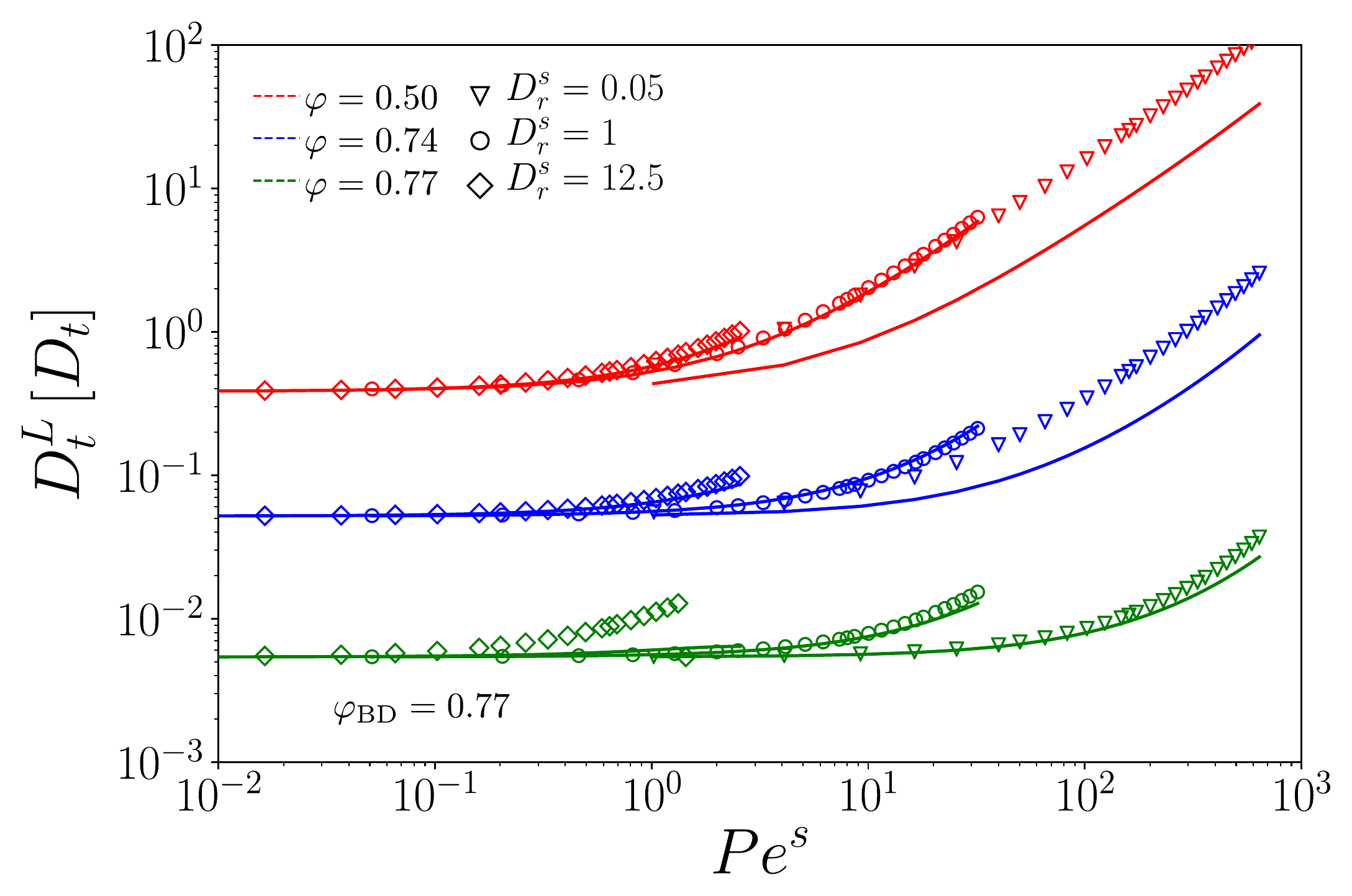}
\caption{\label{fig:active-passive-Dpe-swim}
  Long-time diffusion coefficient $D_t^L(\varphi)$ of an active tracer in
  a passive hard-disk system, as a function of the
  tracer P\'eclet number $\Pe^s$, for the parameters shown
  in Fig.~\protect\ref{fig:active-passive-Dpe}.
  Solid lines repeat the \gls{ABPMCT} results from
  Fig.~\protect\ref{fig:active-passive-Dpe}.
  Symbols show $D_t^L=D_t^{L,0}(1+\Pe^s_\text{eff})$ were
  the effective tracer-P\'eclet number
  $\Pe^s_\text{eff}=(v^s(\varphi))^2/2D_r^sD_t^s$ is evaluated with
  the density-renormalized effective swim velocity $v^s(\varphi)$
  (see text).
}
\end{figure}

A common approach in developing coarse-grained theories of \gls{ABP} is
to account for a density-renormalized swim velocity: Due to interactions,
the average velocity characterizing the particle motion
is no longer the bare self-propulsion speed $v_0^s$ of an individual
\gls{ABP}, but a density-dependent effective swim velocity
$v^s(\varphi)$ \cite{Cates.2015}.
The quantity $v^s(\varphi)$ is in principle a non-equilibrium transport
coefficient onto which the \gls{ITT} framework provides a useful handle.
One can derive, using \gls{ITT}, a generalized Green-Kubo formula for
$v^s(\varphi)$, relating it to the microscopic correlation function of
the particles' orientation-projected forces
\cite{Sharma.2016}.
We have recently obtained a \gls{ABPMCT} expression for the swim velocity
\cite{ReichertPhD} that was shown to be in reasonable qualitative agreement
with simulation data \cite{Reichert.2021}.
Qualitatively, $v^s(\varphi)/v_0^s$ decays from unity at low densities
towards zero at the glass transition, and thus qualitatively explains
the density-dependent reduction of the empirical prefactor $\alpha$
used in Fig.~\ref{fig:active-passive-Dpe}.

A direct comparison of the \gls{ABPMCT} results for the active-tracer
long-time diffusion $D_t^L$ with the expression
$D_t^L\approx D_t^{L,0}(\varphi)(1+\Pe^s_\text{eff}(\varphi))$
(lines and symbols in Fig.~\ref{fig:active-passive-Dpe-swim}) demonstrates
reasonable agreement especially at the highest density studied
($\varphi=0.77$) and not too large $D_r^s$.
Here, $\Pe^s_\text{eff}(\varphi)=v^s(\varphi)^2/2D_r^sD_t^s$ is the
effective P\'eclet number formed with the density-dependent swim speed.
The latter has been evaluated from the theory using the \gls{ITT} expression,
\begin{subequations}
\begin{equation}
  v^s(\varphi)=\frac{v_0^s}{1+\beta\mu/N\int_0^\infty dt\,C(t)}\,,
\end{equation}
employing a \gls{ABPMCT} approximation for the orientation-projected
force autocorrelation function, expressing the latter in terms of a
bilinear functional of the density-correlation functions,
$C(t)\approx\mathcal F^\text{swim}[\tilde\Phi_{ll'}(k,t),\tilde\Phi_{mm'}(k,t)]$
where only terms involving $l,l',m,m'\in\{-1,0,1\}$ enter.
For a derivation and more detailed expressions of the swim-speed functional
we refer to Ref.~\cite{ReichertPhD,Reichert.2021}.
\end{subequations}
Note that the swim velocity depends on $D_r^s$ implicity, however, it
does not capture the complete dependence on rotational diffusion for the
active tracer that is displayed by the results for the long-time
diffusivity in Figs.~\ref{fig:active-passive-Dpe} and \ref{fig:active-passive-Dpe-swim}.

\subsection{Tracer Motion in the Active Bath}

\begin{figure}
\includegraphics[width=\linewidth]{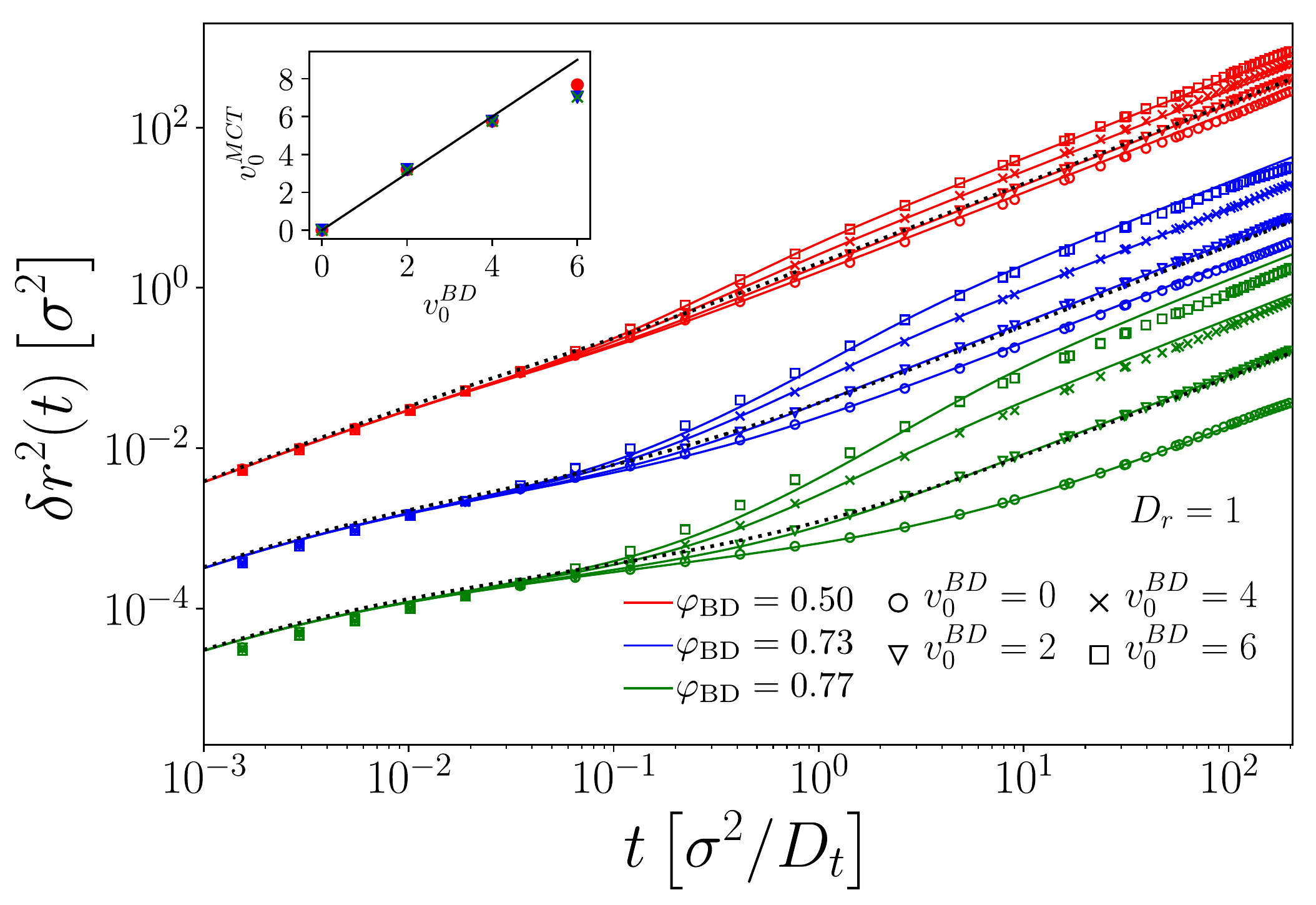}
\caption{\label{fig:passive-active}
  Mean-squared displacements $\delta r^2(t)$ of a single passive
  tracer particle in dense systems of active Brownian hard disks
  at packing fraction $\varphi$ and various self-propulsion velocities
  $v_0$ as indicated. Symbols are results from BD simulations,
  lines are MCT results with an adjustment of the packing fraction
  obtained from the fully passive system, and of the self-propulsion
  velocity (see text and inset).
  For clarity, curve sets for increasing $\varphi$ are shifted down by
  one decade each.
}
\end{figure}

We continue by discussing the tracer motion in an active host
system. One interesting case here is exemplified by the
\gls{MSD} of a passive tracer in a host system of \gls{ABP}
(Fig.~\ref{fig:passive-active}).
Again, for the range of $v_0$ similar to what we discussed in the
reverse case of an active tracer in the passive host, the \gls{MSD}
show a succession of subdiffusive cage motion followed by super-diffusive
escape from the cageing plateau.
It is a clear signature of the non-equilibrium character of the host
system dynamics, that even for the passive tracer, the \gls{MSD} grows
faster than diffusive. This implies that a description of the active
host fluid in an effective-equilibrium framework (such as assigning
an elevated effective temperature to a thermalized fluid with added
activity \cite{Berthier.2013}) cannot capture this dynamics.
Only for the weakly active host system ($v_0=2\,D_t/\sigma$ in
Fig.~\ref{fig:passive-active}) do we find a reasonable description of
the \gls{MSD} in terms of the fully passive-dynamics \gls{MSD} at an
effective reduced packing fraction similar to what was discussed in
connection with Fig.~\ref{fig:active-passive} (dotted lines in
Fig.~\ref{fig:passive-active}).

Qualitatively, the regime of superdiffusion in the case of the
passive tracer, Fig.~\ref{fig:passive-active}, appears more pronounced
the denser is the host system; compare this to the reversed case
of an active tracer in a passive bath, Fig.~\ref{fig:active-passive},
where an increased host-system density serves to more strongly
suppress superdiffusive motion. This is of course intuitive, since
in the latter case, the tracer activity is suppressed by the passive
caging, while in the former case, activity modifies the effective
cage motion that is seen by the tracer.

\Gls{ABPMCT} is again able to account for this nontrivial dynamics
qualitatively. The comparison in Fig.~\ref{fig:passive-active} becomes
even quantitatively satisfactory if one allows for a further empirical
mapping of parameters besides the density mapping that was fixed in the
fully passive system: We find that the influence of host-system activity
on the relaxation dynamics is stronger in the \gls{EDBD} simulations than
it is predicted by \gls{MCT}. The quantitative error can be absorbed in
a rescaling of the self-propulsion velocity $v_0^\text{MCT}$ that enters
the theory calculation. We find reasonable agreement with a linear rescaling,
$v_0^\text{MCT}\approx1.5v_0^\text{BD}$ (inset of Fig.~\ref{fig:passive-active}).

The fact that the effect of the nonequilibrium perturbation on the glassy
dynamics of the host system, in fluidizing that system, is underestimated
by \gls{MCT} is in line qualitatively with previous applications of the
theory to, for example, sheared colloidal suspensions \cite{Fuchs.2009}
or active microrheology \cite{Gazuz.2009,Gazuz.2013,Gruber1,Gruber2}; also there, the introduction of an
empirical scaling factor allowed to bring the theory in quantitative agreement
with simulation data.
In general, one finds that \gls{MCT} overestimates the glassiness of the
relaxation dynamics, and hence it predicts too slow relaxation for a fixed
density $\phi$ and fixed self-propulsion strength $v_0$. Since the effects
of both parameters on the structural relaxation are opposite (increasing
density slows down, increasing activity speeds up the dynamics), it is
plausible that the theory curves for a decreased $\phi$ and an increased
$v_0$ match the simulation data. The mapping of $v_0$ has also been
successful in a description of the relaxation of density fluctuations
at finite $q$ \cite{taggedpaper}.

\begin{figure}
\includegraphics[width=\linewidth]{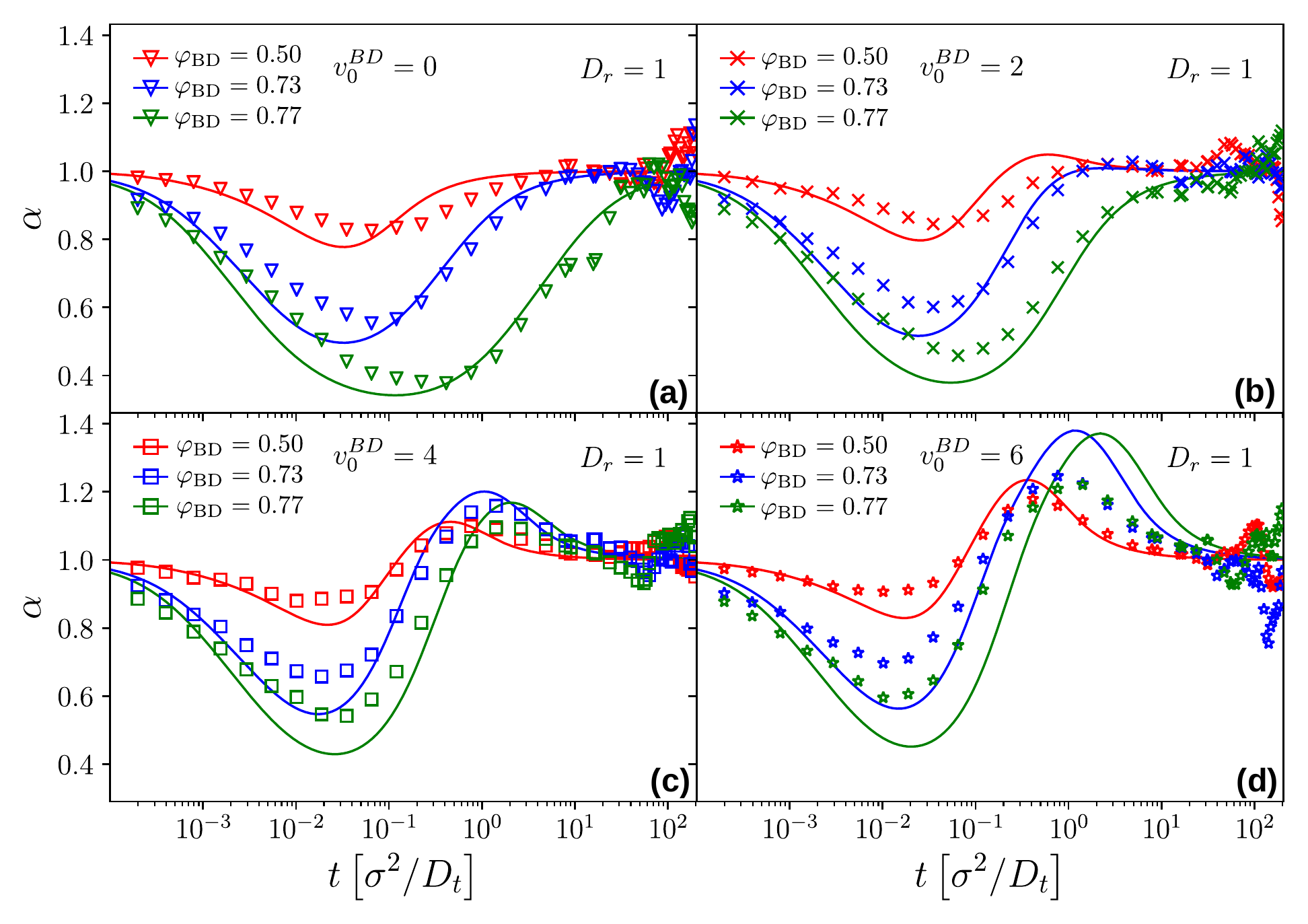}
\caption{\label{fig:passive-active-exp}
  Effective exponents $\alpha(t)=d\log\delta r^2(t)/d\log t$ for the
  MSD of a passive tracer in a host system of active Brownian hard disks,
  corresponding to the data shown in Fig.~\protect\ref{fig:passive-active}.
  Symbols are BD simulation results, lines are from the MCT fits.
}
\end{figure}

As before, an examination of the effective power-law exponent,
$\alpha(t)=d\log\delta r^2(t)/d\log t$ extracted from the logarithmic
derivative of the \gls{MSD}, allows to study in detail the succession
of sub- and super-diffusive regimes (Fig.~\ref{fig:passive-active-exp}).
It becomes apparent that as a general trend, \gls{MCT} overestimates the
extent and strength of both regimes; in particular for the highest
self-propulsion velocity studied here, $v_0=6$, the theory predicts a
pronounced super-diffusive regime around $t=1$ at densities close to the
glass transition; the \gls{EDBD} simulations show superdiffusion to a lesser
extent. This possibly indicates that the simple-minded mapping of
$v_0^\text{BD}$ to an increased $v_0^\text{MCT}$ does not account for
all observations equally well. (We also expect such mapping to only work
in a limit range of $v_0$ and, in particular, $D_r$, but this requires further
investigation.)
Close to $\varphi_c$, the exponents reveal that the \gls{MSD} remains subdiffusive
at all times only for up to $v_0=2\,D_t/\sigma$; this confirms that only
for this weakly active host system, an effective-density passive description
can work.

\begin{figure}
\includegraphics[width=\linewidth]{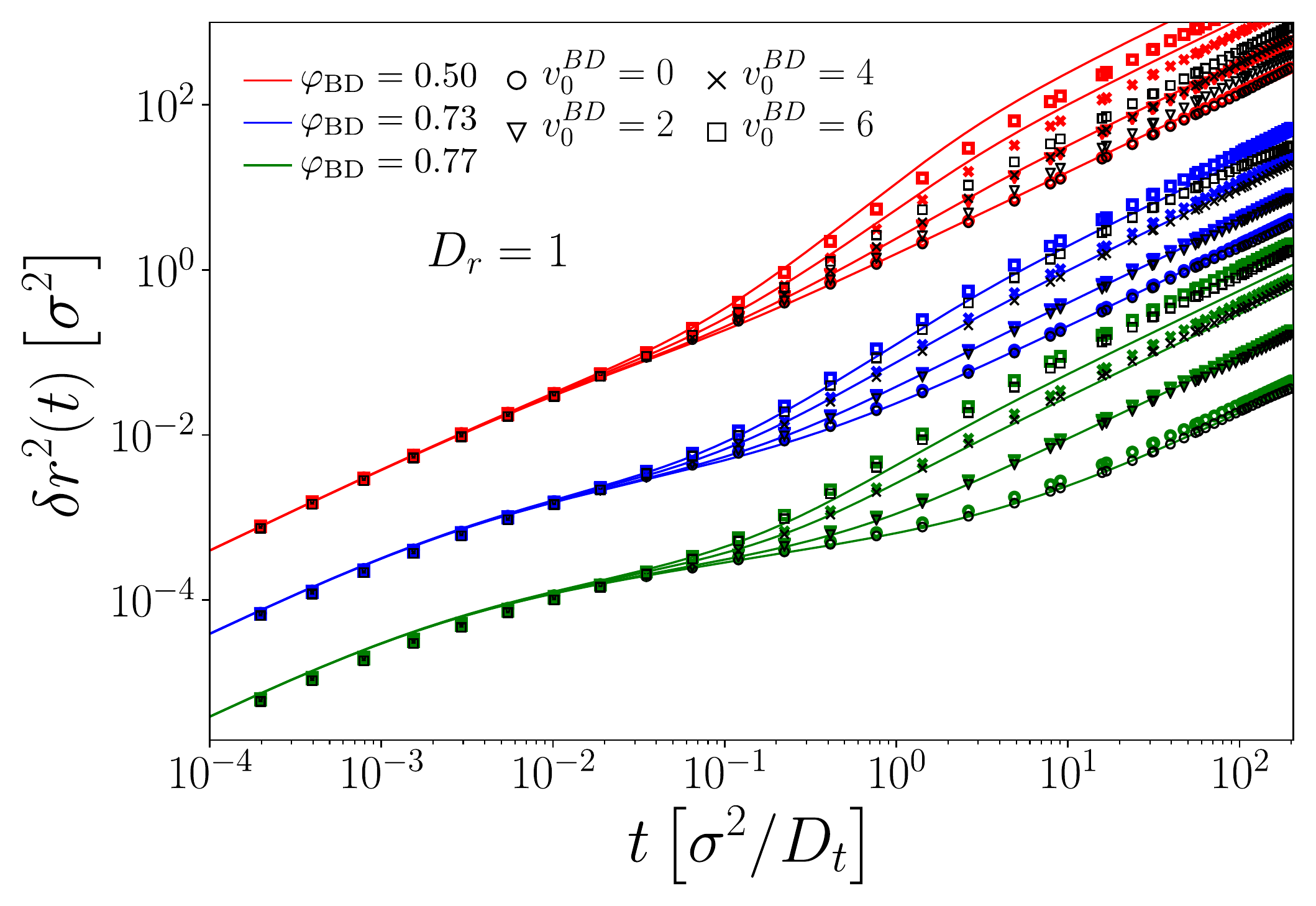}
\caption{\label{fig:active}
  Mean-squared displacements $\delta r^2(t)$
  of an active tracer particle in a host suspension
  of active Brownian disks, at packing fraction $\varphi$ and self-propulsion
  velocities as indicated. Colored symbols are results from BD simulations,
  lines are MCT fits with empirically mapped densities and velocities
  as in Fig.~\protect\ref{fig:passive-active}.
  For clarity, groups of curves corresponding to fixed $\varphi$ are shifted
  downward by one decade each. Black symbols repeat our simulation results
  for a passive tracer in the active host system from
  Fig.~\protect\ref{fig:passive-active}.
}
\end{figure}

Finally, we turn to the \gls{MSD} of an active tracer in a system of
\gls{ABP} (Fig.~\ref{fig:active}).
The results are qualitatively quite similar to the ones
that we have discussed before; intuitively one expects an addition of
the effects discussed in connection with the active tracer in a passive
system, Fig.~\ref{fig:active-passive}, and with the passive tracer in the
active system, Fig.~\ref{fig:passive-active}. Indeed, we observe in the
fully active system (Fig.~\ref{fig:active}) pronounced superdiffusion
succeeding the subdiffusive cage motion at all the densities that are
shown, for sufficiently large $v_0$: at low host system density, it stems from the activity of the tracer
itself, while at high host system density, even the passive tracer acquires
induced superdiffusive motion.
To emphasize the similarity, we compare the \gls{MSD} for the active tracer
in the active host system with those of the passive tracer in that system
(different colored symbols in Fig.~\ref{fig:active}). Indeed, at
the density $\varphi=0.77$ close to the glass transition, both quantities are
nearly identical in our simulations, indicating that here, the dominant
effect comes from the host system activity, and any tracer effectively
follows the collective dynamics. At the lower density $\varphi=0.50$,
the passive tracer shows a far less enhanced superdiffusive regime, since
here the active bath is not yet as effective in transmitting its activity
to the passive tracer.
Note that for sufficiently high density, the \gls{MSD} remains nearly
diffusive and acquires a more pronounced superdiffusive regime when decreasing
the density; this has been also discussed in simulations of
a system of active dumbbells \cite{Mandal.2017}.

The theory correctly captures these two effects: up to the velocity mapping
that is required to quantitatively describe the host system activity,
as described above, both the passive (lines in Fig.~\ref{fig:passive-active})
and the active tracer dynamics (lines in Fig.~\ref{fig:active}) are
quantitatively well described by \gls{ABPMCT}. As anticipated from the
previous discussion, some systematic deviations set in for the largest
$v_0$ that we have studied here. It remains a question for future work
to assess the quality of \gls{ABPMCT} for very large self-propulsion velocities,
once better numerical integration schemes are available for the theory.

\section{Conclusions}
\glsresetall

We have derived equations to describe the \gls{MSD} of active and
passive tracer particles in dense systems of passive or active Brownian
particles. The description is based on the \gls{ITT} framework, a
non-equilibrium statistical physics framework that allows to treat
the activity of \glspl{ABP} as an arbitrarily strong perturbation to
the passive-equilibrium Brownian dynamics. While the formulas,
Eq.~\eqref{eq:msdmct}, are exact in principle, their evaluation
requires knowledge of memory kernels that encode the interaction with the
host particles in terms of positional and dipolar orientational density
fluctuations.

The dynamics at high densities is qualitatively well predicted when the
relevant memory kernels are evaluated using the \gls{ABPMCT},
as our comparison with \gls{EDBD} computer simulations demonstrates.
The good qualitative agreement holds for the range of densities close to
the glass transition and for not too large self-propulsion velocities,
even if the transient correlation functions evaluated within the theory
are compared to the stationary correlations obtained in the simulation.

The most prominent feature of the \gls{MSD} including activity is the
appearance and, at high densities, disappearance of a ballistic regime
of persistent active motion.
For a free \gls{ABP}, superdiffusion appears in a regime set by time and
length scales connected to the self-propulsion velocity and the
reorientational diffusion coefficient. Close to the glass transition,
these time scales compete with the relevant time scales of structural
relaxation, and hence in the \gls{MSD} we observe a typical sequence
of initial passive diffusion, glassy sub-diffusion, followed by
super-diffusive cage breaking at large enough activity, and finally
long-time diffusion. The simulations also demonstrate that for very
large activity, the sub-diffusive cageing regime can be entirely suppressed
by the active motion. These findings are in qualitative agreement with
recent experimental data on the \gls{MSD} of a single active particle
in a colloidal glass former \cite{Lozano.2019}, as we discuss in detail
elsewhere \cite{Reichert.2021}.

In the low-density regime,
absorbing the translational diffusion coefficient $D_t^s$ in the units
of time, the two parameters that quantify active motion, viz.\ its
velocity $v_0^s$ and its persistence time $1/D_r^s$, only enter in
a specific combination through a dimensionless group, the P\'eclet
number $\Pe^s$.
The appearance of a further length scale through the cage effect at
high densities change this, and as a result the motion of the active
tracer depends on both $v_0^s$ and $D_r^s$ separately.

From discussing the various cases of active/passive tracers in active/passive
host systems it emerges that the active motion of the tracer is responsible
for super-diffusive motion as long as the host system is not yet too dense;
in very dense host systems, it is the activity of the host particles that
drive super-diffusive motion even for a passive tracer.

The fact that the extension of \gls{MCT} can describe superdiffusive \gls{MSD} at all is not
trivial. Theories where the angular dynamics is integrated out, and hence
the dynamics of the active particles is mapped onto one described by
an effective Smoluchowksi operator are not a priori able
to capture this.
Especially, for the case of a passive tracer in an active bath, a naive
application of the theory would just assume the standard form of the
passive-\gls{MSD} equations of motion, coupled to enhanced relaxation dynamics in
the bath. Instead, a superdiffusive regime appears in our theory, in good
agreement with simulation.

\section*{Conflicts of interest}

There are no conflicts to declare.

\section*{Acknowledgements}

This project was funded through Deutsche Forschungsgemeinschaft (DFG),
project within the SPP~1726 ``Physics of Microswimmers'', project Vo~1270/7-2.

\begin{appendix}

\section{Features of the Passive-Equilibrium MSD}
\label{sec:eqmsd}

Recall that for a colloidal system in equilibrium, the dynamics is described
by a backward Smoluchowski operator $\smol^\dagger_\text{eq}$
that is self-adjoint in the scalar product
weighted with the equilibrium distribution,
i.e., $\langle f^*\smol_\text{eq}^\dagger g\rangle_\text{eq}=\langle(\smol_\text{eq}^\dagger f^*)g\rangle_\text{eq}$.
The structure of $\smol^\dagger_\text{eq}$ implies $\langle f^*\smol^\dagger_\text{eq} f\rangle_\text{eq}
=-D\langle(\nabla f^*)\nabla f\rangle_\text{eq}=-D\langle|\nabla f|^2\rangle_\text{eq}\le0$ (assuming the diffusion
coefficient to be positive), so that the operator $\smol^\dagger_\text{eq}$ has non-positive
real eigenvalues only.

The density correlation functions are hence completely monotone functions
\cite{Gripenberg,Widder}, i.e., they can be written in the form (specializing to
the tagged-particle correlation function for the sake of the following
argument)
$\phi^s(q,t)=\int\exp[-\gamma t]\,da_q(\gamma)$ with some positive definite
measure $da_q$ that is concentrated on the real axis. For a completely
monotone function, there holds $(-)^k\partial_t^k\phi^s(q,t)\ge0$.

The \gls{MSD} (in $d$ spatial dimensions) follows from
$\delta r^2(t)=\lim_{q\to0}(2d/q^2)(1-\phi^s(q,t))$
and thus $\partial_t\delta r^2(t)$ again is a completely monotone function.
(The \gls{MSD} is thus confirmed to be a monotonically increasing function
of time.)
As a consequence, using the fact that $\delta r^2(t)$ itself is positive,
\begin{equation}
  \frac{\partial\log\delta r^2(t)}{\partial\log t}=\frac{t}{\delta r^2(t)}
  \frac{\partial\delta r^2(t)}{\partial t}\ge 0\,,
\end{equation}
and, since $\delta r^2(0)=0$,
\begin{multline}
  \frac{\partial\log\delta r^2(t)}{\partial\log t}-1
  =\frac{t}{\delta r^2(t)}\left(\frac{\partial\delta r^2(t)}{\partial t}
 -\frac{\delta r^2(t)}{t}\right)\\
  =\frac{t}{\delta r^2(t)}\left(\frac{\partial\delta r^2(t)}{\partial t}
  -\frac1t\int_0^td\tau\,\frac{d}{d\tau}\delta r^2(\tau)\right)\\
  \le\frac{t}{\delta r^2(t)}\left(\frac{\partial\delta r^2(t)}{\partial t}
  -\frac1t\cdot t\frac{\partial\delta r^2(t)}{\partial t}\right)=0\,.
\end{multline}
The latter inequality follows from complete monotonicity:
$\partial_t\delta r^2(t)\ge0$ and $\partial_t^2\delta r^2(t)\le0$ imply
$\partial_\tau\delta r^2(\tau)\ge\partial_t\delta r^2(t)$ for $\tau\le t$,
so that we obtain an upper bound for the integral.
We therefore get
\begin{equation}
  0\le\frac{\partial\log\delta r^2(t)}{\partial\log t}\le 1\,.
\end{equation}
In other words, the effective exponent of the equilibrium Brownian-dynamics
\gls{MSD} is bounded and below unity. Hence, the \gls{MSD} under these
conditions can grow at most diffusively.

\end{appendix}


\bibliography{rsc}
\bibliographystyle{rsc}

\end{document}